# Ultrafast Dynamics for Electron Photodetachment

# from Aqueous Hydroxide. [1]

Robert A. Crowell, [*] Rui Lian, Ilya A. Shkrob, and David M. Bartels [a)]
*Chemistry Division, Argonne National Laboratory, Argonne, IL 60439*
Xiyi Chen and Stephen E. Bradforth
*Department of Chemistry, University of Southern California, Los Angeles, CA 90089*

**Abstract**

Charge-transfer-to-solvent (CTTS) reactions of hydroxide induced by 200 nm monophotonic or 337 nm and 389 nm biphotonic excitation of this anion in aqueous solution have been studied by means of pump-probe ultrafast laser spectroscopy. Transient absorption kinetics of the hydrated electron, $e_{aq}^-$, have been observed, from a few hundred femtoseconds out to 600 ps, and studied as function of hydroxide concentration and temperature. The geminate decay kinetics are bimodal, with a fast exponential component (ca. 13 ps) and a slower power "tail" due to the diffusional escape of the electrons. For the biphotonic excitation, the extrapolated fraction of escaped electrons is 1.8 times higher than for the monophotonic 200 nm excitation (31% vs. 17.5% at 25 °C, respectively), due to the broadening of the electron distribution. The biphotonic electron detachment is very inefficient; the corresponding absorption coefficient at 400 nm is < 4 cm TW$^{-1}$ M$^{-1}$ (assuming unity quantum efficiency for the photodetachment).

For [OH$^-$] between 10 mM and 10 M, almost no concentration dependence of the time profiles of solvated electron kinetics was observed. At higher temperature, the escape fraction of the electrons increases with a slope of $3\times10^{-3}$ K$^{-1}$ and the recombination and diffusion-controlled dissociation of the close pairs become faster. Activation energies of 8.3 and 22.3 kJ/mol for these two processes were obtained. The semianalytical theory of Shushin for diffusion controlled reactions in the central force field was used to model the geminate dynamics. The implications of these results for photoionization of water are discussed.

___________________________________________________________

* To whom correspondence should be addressed: *Tel* 630-252-8089, *FAX* 630-2524993, *e-mail:* rob_crowell@anl.gov.
a) Present address: Radiation Laboratory, University of Notre Dame, Notre Dame, Indiana 46556; *e-mail:* bartels.5@nd.edu.

## I. INTRODUCTION

In liquids consisting of polar molecules with no electron affinity, such as water, alcohols, and acetonitrile, the loss of an electron from a photoexcited anion frequently involves a short-lived mediating state that is unique to these polar liquids. [1] In this charge-transfer-to-solvent (CTTS) state, the excess charge resides in a diffuse orbital that protrudes from the anion into a cavity consisting of solvent molecules that belong to the $1^{st}$ and $2^{nd}$ solvation shells of the parent anion. [2] For anions where the promoted electron originates from a *p*-orbital, such as the halides and hydroxide anions, this diffuse orbital has primarily *s*-like character. This state is stabilized by the pre-existing orientation of the solvent dipoles that leads to an attractive potential for the electron. Similar electrostatic attraction is known to stabilize solvated electrons in the bulk liquid. A fully detached, thermally-relaxed electron is formed as the CTTS state dissociates (150-300 fs) and the resulting species equilibrates with the solvent (< 5 ps). [3,4] Bradforth and coworkers have shown that a large fraction of these solvated electrons (that are spectrally indistinguishable from the bulk species) reside in close proximity to their geminate partner [5,6,7] and weakly interact with it by means of an attractive potential (the energy for this interaction is < 2-3 $kT$). [4] It is presently uncertain what fraction of these close pairs are in actual contact or separated by a solvent molecule, or what is the barrier between these forms. Photodetachment from alkalide anions in tetrahydrofuran also leads to the formation of close pairs; Schwartz and coworkers have interpreted their observations in terms of populations of contact and solvent separated pairs. [8,9] It is likely that (i) both types of these pairs are generated in the photoexcitation of small inorganic anions and (ii) for some of these anions (e.g., alkalides), the corresponding branching ratio depends upon the excitation energy. [10,11]

One of the less studied CTTS reactions is photoinduced detachment of the electron from aqueous hydroxide anion. This photoreaction yields a geminate pair of hydroxyl radical (OH) and hydrated electron ($e_{aq}^-$)

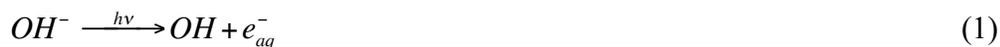

$$OH^- \xrightarrow{h\nu} OH + e_{aq}^- \qquad (1)$$

Although the structure of the aqueous hydroxide is commonly written as OH⁻, it is in fact strongly bonded on average to 3-to-4 water molecules through the oxygen atom leading to strong solvent ordering. [12-15] There is also rapid proton transfer along this hydrogen bonded network; the structure and dynamics of aqueous hydroxide solutions is a topic of active current research. [13,14,15]

The species formed in rxn. (1) rapidly recombine, both geminately and in the bulk, to reform the parent anion

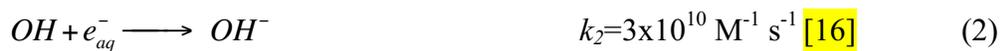

$$OH + e_{aq}^- \longrightarrow OH^- \qquad k_2 = 3 \times 10^{10} \text{ M}^{-1}\text{ s}^{-1} \text{ [16]} \qquad (2)$$

This is one of the fastest reactions for the hydrated electron; [17] still, for $pH > 10$ rxn. (2) competes with deprotonation of the hydroxyl radical: since the $pK_a$ of this radical is 11.54 to 12.1 at 25 °C [18], in alkaline solution, OH rapidly reacts with the hydroxide to yield an O⁻ radical anion:

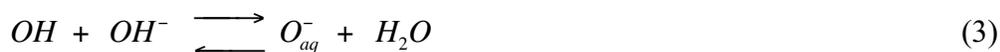

$$OH + OH^- \rightleftharpoons O_{aq}^- + H_2O \qquad (3)$$

Most estimates for rate constant $k_3$ of forward rxn. (3) cluster around $(1.2-1.3) \times 10^{10}$ M⁻¹ s⁻¹ (e.g., ref. [19]). On the other hand, recent kinetic measurement by Hickel, Cortfizen, and Sehested [20] gave $(6.3 \pm 1.3) \times 10^9$ M⁻¹ s⁻¹ for the forward and $10^6$ s⁻¹ for the reverse reaction (previous estimates for the latter, by Zehavi and Rabani [21] and Buxton [19], were as high as $(9.2-9.6) \times 10^7$ s$^{-1}$). The resulting radical anion recombines with the hydrated electron nearly as rapidly as the OH radical:

$$O^-_{aq} + e^-_{aq} + H_2O \longrightarrow 2\ OH^- \qquad k_4 = 2.3 \times 10^{10}\ M^{-1}\ s^{-1}\ [22] \qquad (4)$$

According to Rabani [23], other reactions of OH and O$^-$ with aqueous anions (e.g., halides) and neutral molecules (e.g., alcohols) also have similar rate constants.

Photoreaction (1) satisfies all of the criteria for a CTTS reaction. In the gas phase, the vertical electron detachment energy for OH$^-$ is very low, ca. 1.83 eV [24]. Recent *ab initio* [14] calculations for OH$^-$(H$_2$O)$_n$ clusters (*n<6*) suggest that an addition of several water molecules causes a strong increase in the ionization energy (from 3.62 eV for *n*=1 to 6.2-6.3 eV for *n*=5) and concomitant increase in the CTTS energy (from 3.4 eV for *n*=1 to 5.3-5.6 eV for *n*=5); see Table SII in ref. [14]. The CTTS band of OH$^-$ in the UV has been observed in polar solvents only. In water, this band has a maximum at 6.62 eV (where the extinction coefficient is 3,860 M$^{-1}$ cm$^{-1}$) and a half-width of 0.74 eV; the oscillator strength is 0.109 [25]. The position of the CTTS band maximum for OH$^-$ changes with the solvent polarity in the same manner as that for I$^-$ [1,25]. The only feature that distinguishes OH$^-$ from other CTTS anions is that the UV absorbance does not decrease with increase in the temperature [25]. Apart from this 6.6 eV band, there is some evidence for a low-intensity 7.44 eV band for OD$^-$ in heavy water [25]. This weak VUV band may be due to the second CTTS transition of the hydroxide. Originally, the 6.6 eV band of the aqueous OH$^-$ was classified as a CTTS band because its position compared well with the energy calculated using the "diffuse model" of Stein and Treinin [1,26]. Subsequent time-resolved studies showed that e$_{aq}^-$ is generated shortly after the photoexcitation of OH$^-$ in the 180-200 nm region [27]. Delahay [28,29] and Takahashi et al. [30] showed that in aqueous and acetonitrile solutions, the CTTS energy increases linearly with the threshold electron detachment energy, as measured by photoemission spectroscopy of these solutions. Hydroxide appears on this CTTS energy vs. threshold electron detachment energy plot, between $Cl^-$ and $Br^-$. This correlation further supports the existence of a CTTS state capable of dissociating into OH and $e_{aq}^-$. A quantum yield of 0.11 was given for rxn. (1) in 185 nm [31] and 193 nm [32] photolysis of aqueous $OH^-$ solutions. Femtosecond photodetachment from hydroxide ions was reported by Eisenthal and coworkers, [33] indicating rapid ejection and trapping of electrons. However, only a brief report was given by these workers, and the exact nature of the excited state or the energy deposited in $OH^-$ in this study is unclear.

Our interest in hydroxide is twofold. First, geminate pairs of the OH radical and $e_{aq}^-$ are also produced in the ionization of water. Thus, rxn. (1) provides an opportunity to study the dynamics of such a pair without the interference from hydronium ion ($H_3O^+$) which is also generated in the ionization (e.g., refs. [34-39]). In principle, the contact reaction rate for the hydrated electron with OH can be determined along with its temperature dependence. In addition, there have been suggestions that a short-lived CTTS-like state of the hydroxide (or, perhaps, a contact pair of the OH radical and a pre-thermalized electron) is involved in water photoionization. [39] Because our experiment leads to the unambiguous preparation of a pair, we can put such a suggestion to a rigorous test. Second, $OH^-$ may behave in a qualitatively different way from other CTTS anions. It is a small anion strongly hydrogen-bonded to water, and this bonding may work "against" the formation of a CTTS state if a cavity near the anion is to be filled by the detached electron.

Below, we examine the electron and OH dynamics in 200 nm excitation of aqueous $OH^-$. These dynamics were studied as a function of temperature and hydroxide concentration (section

III.A). Quite different dynamics were observed in 2-photon excitation of OH$^-$ with 337 and 389 nm light (section III.B). Both types of the dynamics are indicative of a weak attraction between the hydrated electron and the OH radical (section IV).

## II. EXPERIMENTAL.

*Laser spectroscopy*. Most of the kinetic measurements reported below were obtained using a 1 kHz Ti:sapphire setup at Argonne. [40] A diode-pumped 5 W Nd:YVO laser was used to pump a Kerr lens mode-locked Ti:sapphire laser operating at 80 MHz (Spectra Physics Tsunami). The 45 fs pulses from the oscillator where stretched to 80 ps in a single grating stretcher. Single pulses were then selected at 1 kHz with a Pockels cell. The 2 nJ pulses were amplified to 4 mJ in a home-built two-stage multipass Ti:sapphire amplifier. The amplified pulses were passed through a grating compressor that yielded Gaussian probe pulses of 60 fs FWHM and 3 mJ centered at 800 nm. The pulse-to-pulse stability was typically 3%.

The amplified beam was split into three parts of approximately equal energy. One beam was used to generate the probe pulses while the other two were used to generate the 200 nm (fourth harmonic) pump pulses. The second beam was passed through two tandem BBO crystals that generated the second (Type I, 500 μm, 29°) and third (Type II, 100 μm, 58°) harmonics. The fourth harmonic is produced by upconversion of the third harmonic (266 nm) and the 800 nm beam in a third BBO crystal (Type I, 100 μm, 68°). Up to 20 μJ of the 200 nm light was produced this way (300-350 fs FWHM pulse). The pump power, before and after the sample, was determined using a calibrated power meter (Ophir Optronics model 2A-SH). The probe pulse was derived from the same Ti:sapphire laser beam and delayed in time as much as 600 ps using a 3-pass 25 mm delay line (a gold corner cube mounted on a Nanomover II translation stage). If not stated otherwise, the probe pulse was 800 nm. To obtain probe light of wavelength other than 800 nm, a white light supercontinuum was generated by focussing the 800 nm light on a 1 mm thick sapphire disk; the probe light was selected using 10 nm FWHM interference filters.

The pump and probe beams were perpendicularly polarized, focused to round spots of 60-150 μm and 20-30 μm in $1/e$ radius, respectively, and overlapped in the sample at 6.5$^\circ$. The spot sizes were determined by scanning the beam with a 10-20 μm pinhole. The maximum fluences of the 200 nm and 400 photons were $2.5 \times 10^{-2}$ and 0.4 J/cm$^2$, respectively. The signal and the reference signals were detected with fast silicon photodiodes, amplified, and sampled using home-built sample-and-hold electronics and a 16-bit A/D converter. A mechanical chopper locked at 50% repetition rate of the laser was used to block the pump pulses on alternative shots. The standard deviation for a pump-probe measurement was typically 10 μOD, and the "noise" was dominated by the variation of the pump intensity and flow instabilities in the jet. The vertical bars shown in the kinetics (e.g., Fig. 1(a)) represent 95% confidence limits for each data point. Typically, 150-200 points acquired on a quasi-logarithmic grid were used to obtain these kinetics.

The experiments at USC were carried out using a 200 kHz Ti:sapphire amplified laser and a pump-probe setup which has been fully described elsewhere. [3,4] Transient absorption signals as low as 10 μOD can be studied using this pump-probe spectrometer. Excitation pulses at 337 nm were generated by frequency doubling the red output of a 389 nm pumped OPA. Pump irradiance at the sample of such pulses was ca. 7 GW/cm$^2$. Alternatively, 389 nm pulses were obtained by frequency doubling the regenerative amplifier fundamental. Pump irradiances up to 175 GW/cm$^2$ can be produced at this wavelength. The 650 nm probe was selected from the white light supercontinuum generated in sapphire with a 25 nm FWHM bandpass interference filter. The level of transient absorption arising from the solvent alone was checked at 389 nm. The contribution is less than 7 %; it is slightly higher at 337 nm. The instrument response function for both spectrometers is 350 fs or better.

*Materials and the flow system.* Hydroxide solutions of low concentration were prepared by mixing 0.989 N KOH volumetric standard (Aldrich) with deaerated nanopure water; concentrated alkaline solutions were prepared by dissolving solid KOH and NaOH pellets (Aldrich, highest purity) in this stock solution. The density and viscosity of concentrated alkaline solutions were taken from ref. [41]. The experiments at the USC were carried out using a 1 mm optical path flow cell with calcium fluoride windows. The experiments at Argonne were carried out with a 150 μm thick, 6 mm wide high-speed (10 m/s) jet with a stainless steel nozzle. To prevent the absorption of $CO_2$ from the air, the jet was enclosed in a box that was constantly purged with dry nitrogen; the solution was also constantly bubbled with $N_2$. The pump and probe beams entered through small holes made in this enclosure. The whole jet assembly was mounted on a 3D translation stage.

An all 316 stainless steel and Teflon flow system was used to pump the solution through this jet. For experiments above room temperature, a 0.5 L reservoir filled with the liquid sample was placed in a circulation bath and equipped with an air-cooled condenser that was used to trap the water vapor carried out by the flow of nitrogen used to purge the alkaline solution. The photolyzed solution returns to the reservoir through this condenser. A gear pump was used to circulate the solution at 0.5-0.75 dm$^3$/min. A high-quality pressure regulator and pulse dampener were used to stabilize the flow and a 60 μm pore sintered steel filter was used to remove cavitation bubbles. Before entering the jet assembly, the solution passed through a heat exchanger. The temperature of the sample was measured using a hair-thin T-type thermocouple inserted 2 cm upstream of the jet nozzle; the temperature of the sample was within 3 $^o$C of the set point of the bath.

**III. RESULTS.**

**A. Monophotonic 200 nm excitation of OH$^-$.**

Fig. 1(a) *(solid line)* shows typical decay kinetics of 800 nm absorbance from hydrated electron obtained in 200 nm excitation of N$_2$-saturated 56 mM KOH in H$_2$O at 25 $^o$C (OH and O$^-$ radicals do not absorb above 300 nm). Under these excitation conditions, almost all of the UV light is absorbed in the 150 μm jet: given the extinction coefficient of 1,000 cm$^{-1}$ M$^{-1}$ for OH$^-$, the 1/*e* penetration depth of the pump light is 78 μm. Exactly the same kinetics were obtained for sodium hydroxide (not shown) and for different concentrations of KOH. Fig. 1(a) shows the comparison between the kinetics obtained for [OH$^-$]=9.8, 56, and 98 mM. At very low hydroxide concentration, [OH$^-$] < 1 mM, and relatively high fluence of the 200 nm light, (0.5-2)x10$^{-2}$ J/cm$^2$, two-photon ionization of water starts to compete with one-photon excitation of OH$^-$ and the kinetics change slightly; for that reason, all of our data were obtained for *pH*>12.

Qualitatively, the kinetics (for convenience, given on the logarithmic time scale) exhibit three time regimes. In the first few picoseconds, there is a fast evolution of the photoinduced absorbance due to the rapid electron thermalization. In Fig. 1(b), we show that after normalization of the kinetics at 5 ps (a delay time where we can assume that the thermalization is over), the signals obtained at 650, 800, and 1000 nm follow each other after the first 5 ps, within the confidence limits. The traces differ in the shape of the short-lived thermalization stage only; this type of spectral evolution in the first 2-3 ps is obtained in all experiments on ionization of neat water [5,38,42,43,44] and has been observed in the CTTS photodetachment from aqueous halides. [5] In the rest of the paper we will focus on the *geminate* dynamics of the hydrated electron observed on a picosecond time scale after the completion of the thermalization phase.

These geminate dynamics exhibit two regimes: (i) a fast exponential decay that occurs on the time scale of tens of picoseconds and (ii) a slower decay that occurs on sub-nanosecond time scale. At 25 $^{o}$C, the overall time profile can be simulated using a biexponential law (e.g., Fig. 2(a)). However, high-temperature data given below in Fig. 3 indicate that the "tail" is actually given by the power law, $t^{-v}$, typical of diffusion-controlled dynamics (see also Fig. 1S; hereafter, extension "S" refers to the figures placed in the Supporting Information). Similar bimodal kinetics (the initial fast exponential decay succeeded by a much slower diffusional decay) were observed in several other CTTS photosystems, such as halide anions in water [3-7] and sodide anions in ethers [8,9,11].

For pump fluences available with our experimental setup ($< 2.5 \times 10^{-2}$ J/cm$^2$), the geminate kinetics showed no dependence on the pump power (see the comparison shown in Fig. 2(a)). The photoinduced 800 nm absorbance observed 10 ps after the 200 nm pulse linearly increased with the photon fluence. The photometric measurement of the quantum yield (QY) of rxn. (1) details of which are given elsewhere [45] gave 0.342±0.08 at $t=5$ ps; this estimate agrees well with the estimates obtained by ultrafast actinometry, assuming unity QY for electron detachment in 200 nm photoexcitation of aqueous ferrocyanide [46]. Using the biexponential fit shown in Fig. 2(a), we estimate that the prompt electron yield (for $t \to 0$) in photoreaction (1) is ca. 0.38 and the free electron yield (for $t \to \infty$) is ca. 0.09. The latter estimate is close to 0.11 obtained by Dainton and Fowles [31] for 185 nm photoexcitation and Iwata et al. [32] and Sauer et al. [47] for 193 nm photoexcitation. The prompt electron yield of 0.38 for 200 nm photoexcitation of $OH^-$ compares well with such yields for other pseudohalides (cf. 0.33 for $HS^-$ and 0.30 for $CNS^-$ [45]); note that these QY's are considerably lower than the near unity values for halides (0.9 for $I^-$ and 0.91 for $Br^-$ [45,48]).

Figs. 3(a) and 3(b) show a family of 200 nm pump – 800 nm probe kinetics obtained for 56 mM KOH solution at several temperatures ranging from 8 to 97 °C. Since the fast decay phase due to thermalization of the hydrated electron becomes less pronounced at higher temperature, these kinetics were normalized at 5 ps, when this thermalization is over. It is clear that the escape fraction of the electron systematically increases with the temperature (Fig. 3(b)). Fig. 3(c) shows that the ratio of the optical densities at 550 ps and 5 ps linearly increases with the temperature. A linear increase in the QY of free electron (estimated from photoproduct analysis) with temperature has been observed by Jortner and coworkers for 254 nm photolysis of iodide [49]. Our ultrafast kinetic measurement suggests that this increase is primarily due to an increase in the survival probability of the geminate pairs.

Remarkably, the time profile of the "tail" does not change with the temperature: In Fig. 1S the plateau value attained at the longest delay time was subtracted from the kinetic traces and the resulting difference traces were normalized at 50 ps. Regardless of the sample temperature, all six kinetics exhibit the same asymptotic behavior. This behavior is expected from the model discussed in section IV.A.

The kinetics shown in Figs. 1, 2 and 3 were obtained at [OH⁻] that was sufficiently low so that the transformation of OH radical into O⁻ occurred after the first 600 ps shown in the kinetic plots. Surprisingly, much the same (normalized) kinetics were obtained in the concentrated alkaline solutions for which rxn. (3) was much faster (Fig. 4). As the concentration of OH⁻ increases into the molar range, the penetration depth of the 200 nm light becomes very short: ca. 1 μm for 5 M solution – and the signal-to-noise suffers due to instability in the jet surface. Still, within the confidence limits of our kinetic measurements (Fig. 4(a)), the normalized traces obtained in 0.1, 1, 4.8, and 9.74 M solutions of KOH look the same, except for small changes in the thermalization dynamics of the electron. This insensitivity to concentration (or ionic strength) will be considered in the Discussion (section IV.B).

**B. Biphotonic excitation of OH⁻.**

In this section we demonstrate that 1 x 200 nm and 2 x 389 nm photoexcitation of aqueous OH⁻ yields qualitatively different electron dynamics despite the fact that the total excitation energies are almost the same.

Since water itself can be ionized by simultaneous absorption of three 400 nm photons [34,50], 2-photon excitation of OH⁻ always competes with this ionization. We refer the reader to the Appendix and Figs. 2S, 3S, and 4S in the Supporting Information for more detail on the photophysics of this competition. As shown therein, provided that the 2-photon quantum yield for electron photodetachment from OH⁻ is unity, the 2 x 400 nm molar absorption coefficient for OH⁻ is very low, ca. 3.8 cm TW$^{-1}$ M$^{-1}$. A simple calculation shows that to observe the 2-photon process exclusively, the pump irradiance should be < 50 GW/cm² (for 2 M OH⁻ solution). At that low irradiance, photoinduced electron absorbance in a 150 μm jet is less than 10 μOD. As it was not possible to study the kinetic behavior for such a small signal using the 1 KHz Ti:sapphire system at Argonne, these measurements were carried out with the USC system, using a 1 mm optical path flow cell.

It was determined that the 650 nm transient absorption signal increased roughly as the square of the 389 nm light irradiance (Fig. 5(c)). At the lower end of the available range, the kinetic profiles were independent of the photon flux. Therefore, these kinetics resulted from *biphotonic* excitation of hydroxide anion in water. These 389 nm traces notably differed from the 400 nm traces obtained at the higher pump irradiance (> 200 GW/cm$^2$) at Argonne. In particular, the latter always showed a contribution from the diffuse geminate pairs generated via water photoionization. This contribution is easily discernable on the semilogarithmic plots shown in Fig. 4S(b).

Fig. 5(a) shows a comparison between the 200 nm (1-photon) and the 389 nm (2-photon) kinetics for OH$^-$. To facilitate the comparison, the kinetic traces were normalized at 5 ps. The solid lines are biexponential least-squares fits in which the time constants for both of the components were the same, but the relative weight of these two components varied. This comparison suggests that (i) the same fast and slow components are observed in both types of photoexcitation, albeit in a different ratio, and (ii) the main difference between these kinetics is a higher fraction of escaped electrons for the 2-photon excitation. As argued in section IV.B, these trends can be rationalized assuming that the initial radial distribution of the electrons in the 2-photon excitation is relatively broad.

In addition to the 389 nm pump - 650 nm probe experiment examined above, a 337 nm pump - 650 nm probe experiment on 1 M OH$^-$ was carried out at USC (Fig. 5(b)). After normalization at 5 ps, these kinetics look identical to the 2 x 389 nm kinetics shown in Fig. 5(b), within the experimental scatter. The small difference could be due to the residual signal from $e_{aq}^-$ generated by water photoionization. At 337 nm, the biphotonic photoionization of water is still inefficient (the onset of the action spectrum is placed at 316 nm [37]). The 2-photon coefficient for absorption of 337 nm photons by $OH^-$ must be very low because even for 1 M NaOH solution and irradiance of 7 GW/cm$^2$, 3-photon ionization of water still contributed 5-7% of the electron absorbance at 650 nm. The fraction of escaped electrons generated in the 2 x 389 and 2 x 337 nm photoprocesses (with total energies of 6.2 and 7.4 eV, respectively) are 1.8 times higher than the same fraction in single 200 nm photon excitation.

## IV. DISCUSSION

**A. Geminate dynamics for *(OH:$e_{aq}^-$)* pairs and diffusion-controlled reactions in a potential well.**

The salient feature of monovalent CTTS photosystems is the formation of a geminate pair inside a solvent cage in which the residual radical/atom and the electron weakly attract each other. [7] The formation of such a cage is not prerequisite: it has been observed for halide anions in water [3-7] and sodide anions in ethers [8,9], whereas electron photodetachment from polyvalent anions, such as ferrocyanide, sulfite, and sulfate does not seem to involve cage formation, but rather long range ejection of the electron. [45,46] The fast exponential regime observed in the geminate dynamics of "caged" pairs corresponds to their escape from and recombination in the cage, whereas the power "tail" is due to slow re-encounters of the electrons that escape from the cage and their geminate partners. [4] Such a picture broadly agrees with optical spectroscopy data and quantum molecular dynamics simulations. [2,51,52,53]

Consider a solvated electron diffusing in a radially-symmetric potential field $U(r)=kT\, u(r)$ and recombining with its geminate partner at reaction radius $r=d$. The mean force potential $U(r)$ mimics the attractive and repulsive forces that exist between the electron and its geminate partner. Such a field can be directly obtained from quantum mechanical - molecular dynamics models of electron photodetachment. [53] In the following, we will assume that the concentration of the excitation events is sufficiently low that cross-recombination of species that belong to different pairs can be neglected. In such a case, the Laplace transform $\Omega(s)$ of the survival probability $\Omega(t)$ for the geminate pair is given by

$$s\Omega(s) = 1 - J(d;s) \qquad (5)$$

where $J(d;s)$ is the Laplace transform of the diffusional flux through the reaction sphere at $r=d$ (in the following, we will assume radial symmetry). For free diffusion ($U(r)=0$) and the absorbing boundary condition at the reaction radius

$$J(d;s) = \frac{d}{r_i}\exp[-k(r_i - d)], \qquad (6)$$

where $k = \sqrt{s/D}$, $D$ is the sum of diffusion coefficients for the geminate partners, and $r_i > d$ is the initial separation between these partners (eq. (7) should be averaged over a distribution $P(r_i)$ of these initial distances). The inverse Laplace transform of eqs. (5) and (6) gives

$$\Omega(t) = 1 - (d/r_i)\, erfc[(r_i - d)/2\sqrt{Dt}] \qquad (7)$$

It follows from this equation that the fraction

$$\Omega_\infty = 1 - d\langle r_i^{-1}\rangle \qquad (8)$$

of electron escape is, in the general case, a function of the initial distribution $P(r_i)$ of the electrons around its geminate partner. Thus, to account for the observed temperature dependence of the kinetic profiles, one needs to postulate a temperature-dependent distribution $P(r_i)$. Our analysis, in which trial distributions $P(r_i)$ were generated and $\Omega(t)$ given by eq. (8) compared with the observed kinetic traces, showed that no choice of this distribution can yield the observed recombination dynamics. We conclude, in agreement with previous studies of halide CTTS, [4,6] that free diffusion cannot explain the observed recombination dynamics for electron-radical pairs generated in electron photodetachment from monovalent anions. The fast exponential decay observed within the first few tens of picoseconds after the excitation suggests that there is a relatively strong interaction between the geminate partners at short distances (3-7 Å), whereas pronounced temperature dependence of $\Omega_\infty$ indicates that there is a weak attractive potential at large separation distances, where the diffusion is almost free.

The analysis of non-homogeneous geminate recombination kinetics in more general cases of finite reaction rates at the reaction radius and diffusion within an interaction potential can be tackled through a variety of approaches, including competitive kinetic models [53] involving different spatial sub-populations, [11] analytic and semi-analytic theories [54-57] and full numerical solution of the diffusion equation. [4,58] In each case, the initial spatial distribution of the generated pairs plays a key role. Two of these approaches have so far been investigated for modeling the recombination dynamics in geminate pairs formed in CTTS reactions. Bradforth and coworkers used a full numerical solution of the Smoluchowski equation [58] for electron diffusion in a potential field that allows an arbitrary form for $P(r_i)$ and $U(r)$ as well as a finite contact reaction rate for the reverse electron transfer reaction. Schwartz and coworkers [8,9] (see also ref. [59]) used a two-state model in which the geminate pairs were divided into close and distant ones; the former pairs were assumed to stay together and decay by first order recombination or dissociation on a picosecond time scale, whereas the second type of pairs were assumed to slowly migrate apart on a sub-nanosecond time scale. Schwartz and coworkers identify the close pairs as "contact" or "immediate" pairs in which the excess electron density envelops the geminate partner, and the distant pairs as "solvent-separated" or free pairs in which the electron is shielded from the partner by one or more solvent molecules. These close and distant pairs have the same spectral properties but exhibit different recombination dynamics when the electron is promoted by absorption of a NIR photon. [8] These two models are not mutually exclusive, because the dynamics of distant pairs becomes diffusional at longer delay times. Implicit in the model suggested by Bradforth and coworkers [4] is that the motion of the electron along the reaction coordinate is purely diffusional and stays so to the point when the geminate partners recombine at the reaction radius.

Below we give a model for geminate dynamics in CTTS photosystems that is based on the semianalytical theory for diffusion-controlled reactions in a potential well developed by Shushin. [55,56,57] This model has a number of advantages in that an explicit form for the potential is not required and it includes the diffusion dynamics. In this model, diffusion-driven escape from a potential well is treated as an equilibrium between the population in the well ($d<r<a$, where $a$ is the Onsager radius at which $|u(a)|=1$), in which the attraction is strong, and in the potential "tail" ($r>a$), where this attraction is negligible. Shushin's equations can be derived using either a two-state equilibrium model [55] or a perturbation treatment of the diffusion equation [56]. In both of these treatments, the "hard" recombination radius $r=d$ is interpreted as the location of the potential barrier between the geminate pair and the product of recombination. Assuming that the well is sufficiently deep (that is, at the potential minimum, $r=r_e$, $u_e=-u(r_e)>1$), Shushin showed that

$$J(d;s) \approx W_r \ n(s) \approx W_r/w(s) , \qquad (9)$$

and

$$w(s) = W + k \ l_d W_d + s \tag{10}$$

where

$$n(s) = \int_d^a dr \ 4\pi r^2 \rho(r,s) \tag{11}$$

is the population in the well, $W = W_r + W_d$ is the escape rate, $W_r$ and $W_d$ are the recombination and dissociation rates, respectively, that are given by

$$W_{r,d} = D l_{r,d}/Z \ , \tag{12}$$

where $Z$ is the partition function

$$Z = \int_d^a dr \ \sigma(r), \ l_r = \int_d^{r_e} dr \ \sigma^{-1}(r) \ and \ l_d = a \equiv \int_{r_e}^{\infty} dr \ \sigma^{-1}(r), \tag{13}$$

$\sigma(r) = r^2 \exp[-u(r)]$, and $a$ is a better estimate for the Onsager radius [55,56]. Eqs. (9) and (10) were derived assuming that the initial distribution function $\rho(r=r_i, t=0)$ inside the well $(d<r_i<a)$ *is quasi-equilibrium* ($\rho(r_i,0) \propto \exp[-u(r_i)]$). If the population relaxation in the well were governed by the diffusion of thermalized electrons, this relaxation would occur in $a^2/D$ [55], i.e., in a few tens of picoseconds. We will assume, however, that the quasi-equilibrium is reached in a few picoseconds, during the electron thermalization (a similar assumption was made in simulating the electron - iodine recombination numerically, [4] by letting $r_i \approx r_e$, and this assumption is supported by molecular dynamics calculations of Borgis and Staib [52,53]). The inverse Laplace transform of eqs. (6) and (9) gives

$$1 - \Omega(\tau) = (1 - \Omega_\infty) M(\alpha; \tau), \tag{14}$$

where we introduced the dimensionless time $\tau = Wt$,

$$\Omega_\infty = p_d \equiv W_d / W \tag{15}$$

is the escape probability of the electron,

$$\alpha = p_d \sqrt{a^2 W/D}, \tag{16}$$

is a dimensionless parameter, and

$$M(\alpha;\tau) = 1 + \operatorname{Im} \lambda^{-1} \exp(\lambda^2 \tau) \operatorname{erfc}(\lambda\sqrt{\tau})/\operatorname{Im} \lambda, \tag{17}$$

where $\lambda = \alpha/2 - i\sqrt{1-\alpha^2/4}$.

Shushin obtained an expansion of function $M(\alpha;\tau)$ in the powers of a small parameter $\alpha$ [56a]. Unlike eq. (17), this expansion is correct for $\alpha<0.3$ only. It is easy to demonstrate, that for a sufficiently long delay time $t$ (or reduced time $\tau$), $M(\alpha;\tau) \approx 1 - \alpha/\sqrt{\pi\tau}$ (so that $\Omega(t) \approx p_d \{1 + a/\sqrt{\pi t D}\}$); in the opposite limit ($\tau >> 1$), $M(\alpha;\tau) \approx 1 - \exp(-\tau)$ (Fig. 5S). Both of these behaviors are consistent with our experimental observations. The dimensionless parameter $\alpha$ is, therefore, a measure of how rapidly the exponential decay is replaced by the diffusion regime due to slow re-encounters of the geminate partners; this parameter depends only on the shape of the attractive potential $U(r)$.

It can be argued from very general principles [58] that the expansion of $w(s)$ in powers of $\sqrt{s}$ given by eq. (10) is *always* correct for small Laplace parameters $s$, though the actual coefficients may differ between the theories. Importantly, empirical estimates for $W_r$ can be used instead of the estimate given by eqs. (12) and (13), as suggested in ref. [56a]. In particular, eq. (12) corresponds to *diffusive* motion across the reaction barrier, which seems unlikely for the system in question (see below).

For $W_r$ given by eq. (12), dimensionless parameters $p_d = l_d/(l_r + l_d)$ (the escape probability for the pair), $\alpha = l_d^2/(Z[l_d + l_r])^{1/2}$, and $W/D = l_d/Z$ are independent of the diffusion coefficient $D$; these parameters depend only on the shape of the reduced potential $u(r)$ and the well depth. The time profile of the geminate kinetics is uniquely defined by these three model parameters or, - if the diffusion coefficient $D$ is known, - by $W_r$, $W_d$, and $a$. Additionally, the reaction constant $K_\infty$ for recombination in the bulk (extrapolated to $t = \infty$) is given by $K_\infty = 4\pi R_{eff} D$, where

$$R_{eff} = a(1 - p_d) \qquad (18)$$

is the effective reaction radius. [57] At 25 °C, the diffusion coefficients are 4.5x10$^{-5}$ cm$^2$/s for the electron and 2.8x10$^{-5}$ cm$^2$/s for the OH radical, respectively, so that $D$=0.73 Å$^2$/ps and $R_{eff}$=5.43 Å. The least-squares fit of picosecond kinetics shown in Fig. 6(a) gives $p_d$=0.173, $\alpha$=0.381, and $W^{-1}$=12.75 ps (or $W/D$=0.107 Å$^{-2}$). Fig. 3(b) shows the least-squares fits for kinetics obtained at higher temperatures. Fig. 7 shows the optimum model parameters plotted as a function of the temperature. No provision was made to take into account rxn. (3), whose rate increases with the temperature, because the concentration of hydroxide was too low (56 mM) to observe deprotonation of OH on the time scale of 600 ps, even at 97 °C [16].

Assuming that the recombination in the well is driven by diffusion alone (eq. (12)) and taking $d$ from $u(d)=0$, we sought the empirical potential

$$u(r) = \varepsilon_0 \left\{ (R_0/r)^n - (R_0/r)^m \right\} \qquad (19)$$

that gives the best agreement with dimensionless parameters $W/D$, $p_d$, and $\alpha$, that depend only on this model potential. To this end, a genetic algorithm for optimization of the parameters $\varepsilon_0$, $R_0$, $n$ and $m$ in eq. (19) was implemented; the additional constraint was to match the experimental $R_{eff}$. The following optimum parameters were obtained: $\varepsilon_0$=8.34 (which corresponds to a well depth of 2.43 $kT$), $R_0$=2.06 Å (that corresponds to a potential minimum at 2.8 Å), $n$=4.77 (repulsion) and $m$=2.1 (attraction). This set of parameters yields $a$=6.76 Å, $l_r$=32 Å, $p_d$=0.174, $\alpha$=0.378, $Z$=377 Å$^3$, $W/D$=0.103 Å$^{-1}$, and $R_{eff}$=5.6 Å. The effective potential thus obtained (shown in Fig. 6(b)) resembles in the well depth the model potential obtained in numerical simulation for the (I:e$_{aq}^-$) pairs by Kloepfer *et al.* [4] Using this potential, one can inquire whether the same potential form can be used to fit the kinetic data obtained at a higher temperature since $u(r)=U(r)/kT$ can be scaled and thus a "temperature dependence" of dimensionless parameters $D/W$, $p_d$, and $\alpha$ obtained. For the potential shown in Fig. 6(b), a small increase in the absolute temperature (10-40%) results in a linear increase in $p_d$ and $\alpha$ and $W/D$ (Fig. 6S). These trends are in agreement with our observations for the (OH:e$_{aq}^-$) pair (Fig. 7); however, the experimental temperature dependencies for these parameters are much steeper (Fig. 7). Furthermore, we found that potentials (19) which yield the best match to the parameters $W/D$, $p_d$, and $\alpha$ at higher temperature (60-97 $^o$C) are shallow (< $kT$) and diffuse ($m$<1.5). Note that even the estimate of $m \approx$ 2 at 25 $^o$C is unrealistic: the mean force potential is likely to be more compact. Furthermore, the basic premise of Shushin's theory is that the potential depth $u_e$ is greater than $kT$, and the shallow potential means that this theory is inapplicable. Thus, it appears that within the framework of Shushin's theory, no reasonable mean force potential can explain the data for the (OH:e$_{aq}^-$) pair. This conclusion, however, relates to our initial assumption that the reaction rate $W_r$ is given by eq. (12), i.e., this reaction is due to diffusion-controlled crossing over the potential barrier. When this assumption is rejected, self-consistency is regained.

In Fig. 8(b), the two rate constants $W_r$ and $W_d$ are plotted vs. $1/T$. This plot indicates that the activation energy for dissociation from the potential well, 22.3±0.9 kJ/mol, is very close to the activation energy for electron diffusion, 20.5±0.2 kJ/mol (Fig. 7S) [60]. The same point is illustrated by the plot of $W_d/D_e$ as a function of the temperature in Fig. 8(b), where $D_e$ is the diffusion coefficient for hydrated electron [60]. On the other hand, the activation energy for the recombination constant, $W_r$, is much lower, 8.3±1 kJ/mol (which is equivalent to ca. 3 $kT$ at 25 °C). Apparently, while the dissociation of the (OH:$e_{aq}^-$) pair is diffusion controlled, the recombination of "caged" partners is not. The activation energy for $W_r$ compares well with 7.9 kJ/mol obtained for rxn. (2) in the bulk, by Elliot and Ouellette (18-150 °C) [16].

Assuming that the diffusion coefficient $D_e$ of the electron accounts for the largest contribution to $D$ (i.e., neglecting the OH diffusion), the Onsager radius $a$ of the potential can be estimated using eq. (16) (Fig. 8(a)). This radius increases linearly with temperature from ca. 5 Å at 8 °C to ca. 7 Å at 97 °C, suggesting that the potential of the mean force becomes more diffuse at higher temperature. The reaction radius $R_{eff}$ estimated using eq. (18) is given in Fig. 8(a); this radius is independent of temperature since the increase in $a$ is cancelled by the decrease in the probability $1-p_d$ of the reaction. The model, therefore, predicts that bulk rxn. (2) is diffusion controlled, contrary to the results of ref. [16]. It is not presently clear, whether this discrepancy is significant. The estimates by Elliot and Ouellette [16] were obtained from a global fit to a complex kinetic scheme that involved >5 other reactions occurring on the same time scale. On the other hand, our simulations neglected a possible temperature effect on the initial electron distribution, which may be significant.

**B. Concentration effects and CTTS photophysics.**

We turn now to the unanticipated observations made in section III, *viz*. (i) the constancy of the geminate decay kinetics as a function of [OH$^-$] and (ii) the fact that 1- and 2-photon photodetachment yield different decay kinetics at nearly the same total excitation energies.

The constancy of the geminate recombination profile observed with concentration in Fig. 4 is remarkable, because these 4.8 M and 9.74 M OH$^-$ solutions are, respectively 21% and 40%, denser and 1.6 and 3.4 times more viscous than pure water. Based on the Stokes law, one would expect to observe considerably slower electron diffusion and thereby slower geminate recombination of the (OH:e$_{aq}^-$) pair. Another factor expected to reduce the diffusion rate is the drag of the slowly-migrating ionic atmosphere on the rapidly-migrating electron, – a well known effect in electrochemistry that is observed for all mobile ions in less concentrated electrolyte solutions [61,62]. The interaction with this atmosphere also increases charge screening that reduces the attraction between the geminate partners. (Large effects of ionic strength on rate constants for reactions of hydrated electrons with charged species have been observed and accounted for via this screening effect. [62,63] Furthermore, assuming that rxn. (3) occurs with rate constant of 1.3x10$^{10}$ M$^{-1}$ s$^{-1}$, the conversion of OH to O$^-$ in a 1-5 M alkaline solution would occur in 20-90 ps which is well within our observation window of 600 ps. Given that *pK$_a$* for the OH radical is ca. 12 [18], in these concentrated solutions the equilibrium concentration of OH is negligible (and, in any case, the backward reaction is too slow to be observed on the picosecond time scale). [19,20,21] We conclude that despite the ongoing transformation of the (OH:e$_{aq}^-$) pair into the (O$^-$:e$_{aq}^-$) pair, no concomitant change in the electron dynamics is observed.

For photodetachment from iodide, similar effects of both reactivity of the detached parent and the ionic atmosphere are in play. The iodine atom generated from iodide adds to I$^-$ with rate constant of (0.88-1.2)x10$^{10}$ M$^{-1}$ s$^{-1}$. [64] This rate constant is close to that for rxn. (3). The resulting dimer radical anion, I$_2^-$, absorbs strongly near 400 nm and in the near infrared. [65] Its formation has indeed been observed probing at 400 nm with the expected kinetics. [66a] The delayed formation of I$_2^-$ on the subnanosecond time scale should also distort the "tails" of the solvated electron kinetics observed at [I⁻] > 0.1 M. Furthermore, there is no *a priori* reason to believe that (I:$e_{eq}^-$) pairs exhibit the same recombination/dissociation dynamics as ($I_2^-$:$e_{aq}^-$) pairs. Nevertheless, experiments carried out in Bradforth's group show that the electron kinetics do not change when [I⁻] changes from 1 mM to 0.2 M, [6,66b] much like the kinetics for OH⁻ given in section III.A. Particularly as the $I_2^-$ formation appears to follow the expected kinetics, this result is also surprising.

The following qualitative explanation may account for the observed constancy of the kinetic profiles with the alkali concentration. Let us first neglect the ionic strength effects and focus on the chemical transformation of the radical. According to Shushin's theory, such a transformation would be negligible provided that this transformation occurs on the time scale *t* for which *Wt>>1*. In this regime, the geminate partners do not interact and, therefore, even if one of the partners undergoes a transformation this does not change the recombination dynamics. Since typical $W^{-1}$ are 10-15 ps, one needs decamolar concentration of the anions to interfere with the geminate dynamics of the electron inside the potential well. Even at this high concentration, the actual changes for the hydroxide system may be negligible because OH and O⁻ have similar reaction constants, and the escape probabilities for the corresponding pairs must be similar. As for the anticipated ionic strength effect on the electron diffusion, this effect requires that the ionic atmosphere rapidly readjusts itself as the electron migrates. In the Debye-Hückel theory of aqueous electrolytes (applicable to ionic concentrations < 0.1 M), the relaxation time for the ionic atmosphere is *55/[ion]* ps, [62] where *[ion]* is the molar concentration of monovalent ions, i.e., at the upper limit of validity of this theory, it takes the atmosphere ca. 500 ps to relax. [62] This time is too long. Even if the relaxation time continues to decrease with [OH⁻] in concentrated alkaline solutions (and this trend is opposed by the decrease in cation diffusion and ion pairing), it still takes extremely large concentrations of the electrolyte to reduce this relaxation time to the time scales comparable to $W^{-1}$.

As a result of these observations, a detailed study of ionic strength effects on electron radical recombination reactions has been conducted [67]. In addition, separate experiments that tracked the 266 nm absorbance of OH radical as a function of the initial OH$^-$ concentration were carried out. [68] That latter study suggested that forward rxn. (3) slows down considerably in concentrated alkaline solutions. We direct the reader to refs. [67] and [68] for a full discussion of the problems. For our purposes, it is significant that sub-nanosecond recombination chemistry is invariant to [OH$^-$] (and, more generally, the ionic strength of the solution [67]) as it allows comparison of the one- and two- photon initiated detachment reactions.

The first major difference in the 1- and 2-photon excitation of hydroxide is that the 2-photon detachment process appears to be extremely inefficient (section III.B). The 2-photon efficiency for electron photodetachment from OH$^-$ at 400 nm is 35 times lower than that for iodide at the same wavelength (see refs. [69] and [70]). This efficiency factor involves both the two-photon cross-section for absorption and the prompt yield of electrons upon excitation. This large difference may indicate that 2-photon CTTS of OH$^-$ is a forbidden photoprocess. In the gas phase, the ground state OH$^-$ anion has $\Sigma$ representation whereas the OH radical has $\Pi$ representation [24,71]. The first excited electronic state ($A\ ^2\Sigma^+$) of the OH radical can be excluded from consideration because it has > 4 eV higher energy than the ground state [71]. To conserve the projection of the orbital momentum on the OH axis, the electron wave should be *s*- or *d*- for 1-photon detachment and *p*- or *f*- for 2-photon detachment. [72] The role of the electron wave symmetry and orbital momentum conservation in 1-photon CTTS dynamics has been recently studied by Schwartz and coworkers, [73] and we address the reader to their theoretical work for more detail.

Two possible explanations can be given to account for inefficient biphotonic electron detachment from $OH^-$. The choice between these two explanations depends on whether the low efficiency constitutes an anomaly or the rule, and that is not yet known. In the first scenario, the generation of a *p*-wave electron from the $OH^-$ is viewed as a low-probability event because it requires a fortuitous orientation of the OH radical and a figure-eight cavity that stabilizes the outer lobes of the CTTS *p*-function. Since both atoms in the parent $OH^-$ anion are hydrogen bonded to water molecules, it is unlikely that such a "cavity" can form in the direction that is required to compensate for the orbital momentum of the $\Pi$ state. Indeed, molecular dynamics calculations [13], IR-Raman studies of alkaline solutions (see a review in ref. [15]) and $OH^-$($H_2O$)$_n$Ar$_m$ clusters (e.g., [12]) suggest that the O atom is strongly hydrogen bonded to 3-4 water molecules. Such a rigid structure leaves little room for the cavity formation. In this scenario, the low yield of 2-photon CTTS for $OH^-$ is due to strong hydrogen bonding.

An entirely different rationale has been suggested by Schwartz [74]. In that scenario, the orbital symmetry disfavors 2-photon CTTS for *all* nearly-spherical anions, including the $OH^-$ and iodide anions. However, for the latter, spin-orbital interaction relaxes the orbital momentum conservation and makes 2-photon CTTS more favorable. In such a case, heavier, larger anions would possess larger cross sections for electron detachment than smaller, lighter anions. Only further study can establish which one of these factors, hydrogen bonding or spin-orbital interaction, is responsible for the observed difference between the $OH^-$ and the $I^-$. In particular, a systematic study on the dependence of 2-photon cross sections and quantum yields for pseudohalides and halides is needed to obtain the general pattern.

However small is the probability for 2-photon induced electron detachment from OH⁻, when this photoreaction occurs, the initial radial distribution of hydrated electrons may be different from that attained in 1-photon rxn. (1), as these localized species originate from electron waves of different symmetry. Below, we argue that in the framework of Shushin's theory, the difference between the kinetics shown in Fig. 5(a) can be understood in terms of a broader radial distribution of the electrons for *biphotonic* excitation. This distribution extends over the Onsager radius $a$ of the potential thereby increasing the escape fraction of the electrons. Using the equilibrium model of ref. [55], the Laplace transform $n(s)$ of the population of the particles in the well (at $r<a$) is

$$n(s) \approx \left[\Theta(a-r_i) + J(a;s)\ \Theta(r_i-a)\right]/w(s) \qquad (20)$$

where $J(a;s)$ is the free-diffusion flux through the Onsager surface given by eq. (7) and $r_i$ is the initial separation of geminate partners. Eq. (20) should be averaged over the initial electron distribution $P(r_i)$. Then, the escape fraction is given by

$$\Omega_\infty = 1 - (1-p_d)\langle \min(1, a/r_i)\rangle \qquad (21)$$

For an r²-exponential distribution with $\langle r_i \rangle = 3b$,

$$4\pi r_i^2 P(r_i) = (2b^3)^{-1} r_i^2 \exp(-r_i/b), \qquad (22)$$

and the integration can be carried out analytically to obtain

$$n(s) = \left\{1 - \exp(-\upsilon)\left[1 + \upsilon + \upsilon^2/2 - \upsilon/2\ \left(\upsilon(kb+1)+1\right)/(kb+1)^2\right]\right\}/w(s), \qquad (23)$$

so that

$$\Omega_\infty = p_d + (1-p_d)\exp(-\upsilon)[1 + \upsilon/2], \qquad (24)$$

where $\upsilon = a/b$. The inverse Laplace transform of $\Omega(s)$ can be carried out numerically using an FFT algorithm. Fig. 9(b) shows a family of $\Omega(t)$ kinetics plotted as a function of the ratio $\upsilon$. Assuming that for 1 x 200 nm excitation of OH$^-$ $\upsilon >>1$ and $p_d$=0.173, we obtain, using eq. (24) and substituting the experimental estimate $\Omega_\infty \approx 0.31$, that for 389 nm (2-photon) dynamics, $\upsilon \approx$ 2.62 and $\langle r_i \rangle \approx 1.15a$. As shown in Fig. 9(b), a decrease in $\upsilon$ (broader electron distribution) results in kinetics that are similar to trace (ii) in Fig. 5(a): the escape fraction $\Omega_\infty$ increases and the weight of the fast component decreases with the increase in $\upsilon$. The best match between the simulated and experimental kinetic profiles is for $\upsilon \approx 3$, at which $\langle r_i \rangle \approx a$ (Fig. 9(a)). Thus, the biphotonic kinetics can be interpreted in terms of an initial electron distribution with the mean distance equal to the Onsager radius of the mean force potential that is estimated to be 5-6 Å for the given pair. A similar approach has been recently used by us to explain the difference between the recombination kinetics for electron detached from iodide by 200 nm and >225 nm light: both sets of the kinetics can be simulated assuming the same dynamic parameters and mean force potential; the only difference seems to be a broader electron distribution in 200 nm photoexcitation [45]. It is noteworthy that the excitation energy dependence for the geminate kinetics observed by Barthel and Schwartz in electron detachment from sodide in THF [11] can also be interpreted in the same way, as systematic broadening of the electron distribution. It must be stressed that the scenarios considered above are speculative and need to be explored using molecular dynamics and quantum mechanical calculations.

The observations of differing geminate recombination with photoelectron ejection range provide an explanation for the difference between the current CTTS detachment results and those published in earlier work by Eisenthal and coworkers. [33] These authors report rapid generation of $e_{aq}^-$ in 313 nm laser photolysis of aqueous OH$^-$. The excitation was either bi- or tri- photonic, and picosecond kinetics of the electron decay resembled those for photoionization of neat water, with an escape fraction of the electron of 70% attained at a delay time of 70 ps. Although the kinetics reported here for hydroxide when resonantly excited into its CTTS band (section III) look different from that reported in ref. [33], the latter are consistent with preliminary measurements made at USC using 300 nm two-photon excitation of concentrated alkali solution. Once 8.3 eV is deposited into the photosystem, the electron is ejected to a much longer distance and may be directly coupled to the water conduction band. [10,45]

**C. Implications for water photolysis and radiolysis.**

As mentioned in the Introduction, one of the main motivations for studying hydroxide CTTS is the fact that geminate pairs of OH and electron also occur in photo- and radiation induced ionization of liquid water:

$$H_2O \xrightarrow{h\nu} H_3O^+ + HO + e_{aq}^- \qquad (25)$$

According to numerous studies (e.g., refs. [34-39]), geminate decay of the electron generated in rxn. (25) proceeds mainly via. rxn. (2), though electron recombination with the hydronium ion also contributes to the electron decay (ca. 18% of recombination events at 25 °C [38]). The usual approach to the simulation of the ensuing dynamics is to assume free diffusion and to use the so-called independent reaction time (IRT) model. [75] The alternative approach is to perform Monte-Carlo simulations; [75] in any case, no attraction or caging between the OH radical and the electron is assumed.

Our results suggest that this approach may be incorrect since hydroxide CTTS yields a pair whose geminate dynamics indicates such an attraction, and similar dynamics were observed for other electron - radical pairs. [2-11] This weak attraction is not specific to the pairs that are generated via the CTTS process, i.e., it cannot be related to the nature of the solvent cage that is formed in this photoprocess or a "memory" about the water structure around the parent anion. Indeed, the effects of such an interaction can be traced out to 50-100 ps. It seems unlikely that any specific configuration of water molecules around the OH radical or its parent anion would survive for that long period of time. The observation of chemically induced electron spin polarization in the CTTS photosystems by Fessenden and coworkers [76] and Bussandri and van Willigen [77] also suggests that caging occurs even in random encounters of radicals with electrons, as otherwise no spin exchange would be possible in such pairs due to their rapid recombination. [78] While the exact mechanism by which the attraction between the OH radical and $e_{aq}^-$ arises is not presently clear, its very existence contradicts the basic assumptions of the IRT and Monte-Carlo simulations.

Why was this cage formation not observed before? As explained in the previous section, the degree by which the fast exponential component that corresponds to escape from the cage can be discerned from the geminate kinetics depends on the width of the initial electron distribution. For hydroxide CTTS, $\upsilon > 3$ (section IV.B). For 2 x 266 nm or 3 x 400 nm [34] photon ionization of water, $\langle r_i \rangle \approx 9-10$ Å and $\upsilon \approx 1.6$. For ionization using high energy photons and in radiolytic spurs, $\langle r_i \rangle \approx 20-30$ Å [34] and $\upsilon < 0.6$. Examination of Fig. 9(b) shows how difficult it is to discern the exponential component in the last two cases. Additionally, hydrated electron interacts and recombines with the hydronium ion, and that further biases the resulting kinetics towards agreement with the free diffusion prescription.

Recently, Laubereau and coworkers suggested that contact (OH:$e_{aq}^-$) pairs are generated in 2 x 273 nm photoionization of water and asserted that such pairs account for a short-lived (ca. 100 fs) absorbance near the fundamental OH stretch of water [39]. According to their reaction scheme, photoionization of water proceeds through the formation of a hydronium ion and an "electronically excited OH$^-$" anion which rapidly dissociates to a contact (OH:$e_{aq}^-$) pair. The short-lived 2-5 μm absorbance is thought to be from the OH stretch of a hydroxyl radical that interacts with the electron in the same cavity.

Our results do not support such a scenario. "Electronically excited OH$^-$" and its partially dissociated state that are invoked by Laubereau and coworkers [39] would yield a CTTS-like (OH:$e_{aq}^-$) pair with short separation between the OH radical and the electron. Such a pair would exhibit the characteristic time profile of electron decay that was observed in our transient absorption dynamics for hydroxide CTTS. The energetics are also unfavorable. The total excitation energy of 2 x 273 nm photoexcitation is 9.1 eV and > 0.83 eV of this energy is consumed to break the H-O bond of water and solvate the resulting proton. [34] Thus, the OH$^-$ fragment has at most 8.3 eV of the excess energy, which is close to the threshold energy of 8.45 eV for the OH$^-$ photoemission (into the gas phase) [28]. This energy is only 0.9 eV higher than the energy absorbed by OH$^-$ in 2 x 337 nm (7.4 eV) photoexcitation, which yielded the characteristic bimodal kinetics shown in Fig. 5(b). No bound excited states of OH$^-$ in water other than its 6.6 and 7.44 eV CTTS states are known [25].

**V. CONCLUSION**

One- (200 nm) and two- (337 and 389 nm) photon induced electron detachment from aqueous OH$^-$ have been studied on the picosecond time scale. In both of these cases, geminate decay kinetics for the hydrated electron are bimodal, with a fast exponential component (ca. 13 ps) and a slower "tail" due to diffusional escape of the electrons on the sub-nanosecond time scale. For biphotonic excitation, the extrapolated escape fraction of the electron is 1.8 times higher than for monophotonic excitation (31% vs. 17.3% at 25 $^o$C, respectively). We speculate that this difference is due to the broader electron distribution attained in the biphotonic excitation. The prompt quantum yield of the 200 nm photodetachment is ca. 0.38 for hydroxide vs. 0.9 for iodide. [45] The two-photon excitation of CTTS at 400 nm is 35 times less efficient than that of the aqueous iodide.

Picosecond electron dynamics observed in the monophotonic excitation of the CTTS of OH$^-$ were studied as a function of [OH$^-$] and the temperature. The temperature effect (studied over the temperature interval of 8-to-90 $^o$C) is mainly in the increase of the escape fraction of the electrons (ca. 3x10$^{-3}$ per $^o$K) and faster recombination (with activation energy of 8.3 kJ/mol) and diffusion-controlled dissociation (with activation energy of 22.3 kJ/mol) of the geminate pairs in the potential well. The activation energy for recombination of caged e$_{aq}^-$ and OH radical compares well with 7.9 kJ/mol obtained for rxn. (2) in the bulk. No concentration dependence was observed, from 10 mM to 10 M. Neither the ionic strength effects (due to the charge screening and the drag of the ionic atmosphere) nor the effect of ongoing rxn. (3) were observed on the sub-nanosecond time scale. This result is surprising, and is the subject of an extensive parallel study [67,68].

The picosecond kinetics were simulated in a model where the electron migrates by diffusion and interacts with its geminate partner by means of a weak attractive potential. The semianalytical theory of Shushin was used to model the geminate dynamics. These simulations suggest that the mean force potential becomes more diffuse at higher temperature, with the Onsager radius linearly increasing from 5 Å at 8 $^{o}$C to 7 Å at 97 $^{o}$C. This trend is consistent with the notion that the potential is mainly due to H-bonding of water molecules in the solvation shells rather than direct electrostatic interaction of the geminate partners.

Our results suggest that dynamic models for photoionization of water should be revised to include this weak interaction of photogenerated electron with the OH radical. Such an interaction is directly observed in the geminate pairs generated via 1- and 2-photon induced photodetachment from OH$^-$; there is no reason to expect that it is lacking for the same pairs generated via water ionization. On the other hand, our results do not support the recent suggestion by Laubereau and coworkers [39] that *contact* (OH:e$_{aq}^-$) pairs are generated in water photoionization and yield the short-lived absorbance in the 2-5 µm region [39]. This photoinduced change in the IR light transmission originates from species other than the OH radical; perhaps, from the water itself.

## VI. ACKNOWLEDGEMENT.

We thank Prof. B. J. Schwartz and Dr. C. D. Jonah for useful discussions. The research at the ANL was supported by the Office of Science, Division of Chemical Sciences, US-DOE under contract number W-31-109-ENG-38. The research at USC was supported by the National Science Foundation and the David and Lucile Packard Foundation. SEB is a Cotrell Scholar of the Research Corporation.

**References.**

1. M. J. Blandamer and M. F. Fox, Chem. Rev. **70**, 59 (1970).

2. S. E. Bradforth and P. Jungwirth, J. Phys. Chem. A **106**, 1286 (2002).

3. J. A. Kloepfer, V. H. Vilchiz, V. A. Lenchenkov, and S. E. Bradforth, Chem. Phys. Lett. **298**, 120 (1998).

4. J. A. Kloepfer, V. H. Vilchiz, V. A. Lenchenkov, X. Chen, and S. E. Bradforth, J. Chem. Phys. **117**, 776 (2002).

5. V. H. Vilchiz, J. A. Kloepfer, A. C. Germaine, V. A. Lenchenkov, and S. E. Bradforth, J. Phys. Chem. A **105**, 1711 (2001).

6. J. A. Kloepfer, V. H. Vilchiz, V. A. Lenchenkov, A. C. Germaine, and S. E. Bradforth, J. Chem. Phys **113**, 6288 (2000).

7. J. A. Kloepfer, V. H. Vilchiz, V. A. Lenchenkov, and S. E. Bradforth, in *Liquid Dynamics: Experiment, Simulation, and Theory* (2002), Vol. 820, pp. 108.

8. I. B. Martini, E. R. Barthel, and B. J. Schwartz, J. Am. Chem. Soc. **124**, 7622 (2002).

9. E. R. Barthel, I. B. Martini, and B. J. Schwartz, J. Chem. Phys. **112**, 9433 (2000); Science **293**, 462 (2001).

10. X. Chen, J. A. Kloepfer, S. E. Bradforth, R. Lian, R. A. Crowell, and I. A. Shkrob, *(in preparation)*.

11. E. R. Barthel and B. J. Schwartz, Chem. Phys. Lett. **375**, 435 (2003).

12. W. H. Robertson, E. G. Diken, E. A. Price, J.-W. Shin, and M. A. Johnson, Science **299**, 1367 (2003) *and references therein.*

13. M. E. Tuckerman, D. Marx, M. L. Klein, M. Parrinello, Science **275**, 817 (1997), Nature **417**, 926 (2002) *and references therein.*

14. M. Masamura, J. Chem. Phys. **117**, 5257 (2002).

15. H.-K. Nienhuys, A. J. Lock, R. A. van Santen, and H. J. Bakker, J. Chem. Phys. **117**, 8021 (2002) *and references therein.*

16. A. J. Elliot and D. C. Ouellette, J. Chem. Soc. Faraday Trans. **90**, 837 (1994).

17. A. J. Elliot, *Rate Constants and G-Values for the Simulation of the Radiolysis of Light Water over the Range 0-300 $^oC$*, AECL report 11073, COG-95-167 (AECL Research, Chalk River Laboratories, Chalk River, Ontario, Canada, 1994).

18. G. A. Poskrebyshev, P. Neta, and R. E. Huie, J. Phys. Chem. A **106**, 11488 (2002); J. Rabani and M. S. Matheson, J. Phys. Chem. **70**, 761 (1966); J. L. Weeks and J. rabani, J. Phys. Chem. **70**, 2100 (1966); F. S. Dainton and D. C. Walker, Proc. Roy. Soc. A **285**, 339 (1965).

19. G. V. Buxton, Trans. Farad. Soc. **66**, 1656 (1970).

20. B. Hickel, H. Corfitzen, and K. Sehested, J. Phys. Chem. **100**, 17186 (1996).

21. D. Zehavi and J. Rabani, J. Phys. Chem. **75**, 1738 (1971).

22. J. Rabani and M. S. Matheson, J. Am. Chem. Soc. **86**, 3175 (1964).

23. J. Rabani, *Radiation Chemistry of Aqueous Systems*, ed. G. Stein (Weizmann Science Press, Jerusalem, 1968), p. 229.

24. H. Hotop, T. A. Patterson, and W. C. Lineberger, J. Chem. Phys. **60**, 1806 (1974); P. A. Schulz, R. D. Mead, P. L. Jones, and W. C. Lineberger, J. Chem. Phys. **77**, 1153 (1982).

25. M. Fox, R. McIntyre, and E. Hayon, Faraday Discuss. Chem. Soc. **64**, 167 (1977); molar absorptivity of aqueous hydroxide can be found in S. O. Nielsen, B. D. Michael, and E. J. Hart, J. Phys. Chem. **80**, 2482 (1976); less reliable estimates are given by J. Jortner, B. Raz, and G. Stein, J. Chem. Phys. **34**, 1455 (1961) and J. L. Weeks, G. M. A. C. Meaburn and S. Gordon, Rad. Res. **19**, 559 (1963).

26. G. Stein and A. Treinin, Trans. Farad. Soc. **55**, 1086 (1959).

27. M. S. Matheson, W. A. Mulac, and J. Rabani, J. Phys. Chem. **67**, 2613 (1963).

28. P. Delahay, *Electron Spectroscopy: Theory, Techniques and Applications*, v. 5; ed. C. R. Brundle and A. D. Baker (Academic Press, New York, 1984), p. 124; P. Delahay, Chem. Phys. Lett. **89**, 149 (1982).

29. P. Delahay and K. Von Burg, Chem. Phys. Lett. **83**, 250 (1981); I. Watanabe, J. B. Flanagan, and P. Delahay, J. Chem. Phys. **73**, 2057 (1980).

30. N. Takahashi, K. Sakai, H. Tanida, and I. Watanabe, Chem. Phys. Lett. **246**, 183 (1995); K. Sakai, I. Watanabe, and Yu. Yokoyama, Bull. Chem. Soc. Jpn. **67**, 360 (1994).

31. F. S. Dainton and P. Fowles, Proc. Roy. Soc. A **287**, 295 and 312 (1965); see also J. Barret, M. F. Fox, and A. L. Mansell, J. Phys. Chem. **69**, 2996 (1965).

32. A. Iwata, N. Nakashima, M. Kusaba, Y. Izawa, and C. Yamanaka, Chem. Phys. Lett. **207**, 137 (1993).

33. F. H. Long, H. Lu, X. Shi, and K. B. Eisenthal, Chem. Phys. Lett. **169**, 165 (1990); J. Chem. Phys. **91**, 4413 (1989).

34. R. A. Crowell and D. M. Bartels, J. Phys. Chem. **100**, 17940 (1996).

35. M. U. Sander, K. Luther, J. Troe, J. Phys. Chem. **97**, 11489 (1993).

36. J. L. McGowen, H. M. Ajo, J. Z. Zhang, B. J. Schwartz, Chem. Phys. Lett. **231**, 504 (1994).

37. C. L. Thomsen, D. Madsen, S. R. Keiding, J. ThØgersen, and O. Christiansen, J. Chem. Phys. **110**, 3453 (1999).

38. D. Madsen, C. L. Thomsen, J. ThØgersen, and S. R. Keiding, J. Chem. Phys. **113**, 1126 (2000).

39. R. Laenen and T. Roth, J. Mol. Struct. **598**, 37 (2001); R. Laenen, T. Roth, and A. Laubereau, Phys. Rev. Lett. **85**, 50 (2000).

40. R. A. Crowell, R. Lian, M. C. Sauer, Jr., D. A. Oulianov, and I. A. Shkrob, Chem. Phys. Lett. **383**, 481 (2004).

41. P. M. Sipos, G. Hefter, and P. M. May, J. Chem. Eng. Data **45**, 613 (2000); T. M. Herrington, A. D. Pethybridge, and M. G. Roffey, J. Chem. Eng. Data **31**, 31 (1986).

42. R. A. Crowell, R. Lian, and I. A. Shkrob, *in preparation.*

43. C. Pepin, T. Goulet, D. Houde, and J.-P. Jay-Gerin, J. Phys. Chem. A **101**, 4351 (1997).

44. A. Hertwig, H. Hippler, A. N. Unterreiner, and P. Vohringer, Ber. Bunsenges. Phys. Chem. **102**, 805 (1998); J. Phys.: Condens. Matter **12**, A165 (2000); Phys. Chem. Chem. Phys. 4, 4412 (2002).

45. R. Lian, D. A. Oulianov, R. A. Crowell, I. A. Shkrob, X. Chen and S. E. Bradforth, *to be submitted to J. Phys. Chem. A.*

46. V. A. Lenchenkov, J. A. Kloepfer, V. H. Vilchiz, and S. E. Bradforth, Chem. Phys. Lett. **342**, 277 (2001).

47. M. C. Sauer, Jr., R. A. Crowell, and I. A. Shkrob, *to be submitted to J. Phys. Chem. A.*

48. The prompt quantum yield for electron detachment from halide has been estimated indirectly, by Kloepher et al. (ref. [4]) using actinometric measurement by Iwata et al. (ref. [32]); we have recently reaffirmed this estimate by direct ultrafast photometry, ref. [35].

49. J. Jortner, M. Ottolenghi, and G. Stein, J. Phys. Chem. **68**, 247 (1964).

50. R. Naskrecki, M. Ménard, P. van der Meulen, G. Vigneron, and S. Pommeret, Opt. Comm. **153**, 32 (1998).

51. W.-S. Sheu and P. J. Rossky, Chem. Phys. Lett. **202**, 186 and 233 (1993); J. Am. Chem. Soc. **115**, 7729 (1993); J. Phys. Chem. **100**, 1295 (1996).

52. D. Borgis and A. Staib, Chem. Phys. Lett. **230**, 405 (1994); J. Chim. Phys. **93**, 1628 (1996); J. Phys.: Condens. Matter **8**, 9389 (1996); J. Mol. Struct. **436**, 537 (1997); J. Chem. Phys. **103**, 2642 (1995)

53. D. Borgis and A. Staib, J. Chem. Phys. **104**, 4776 and 9027 (1996).

54. H. Sano and M. Tachiya, J. Chem. Phys. **71**, 1276 (1979).

55. A. I. Shushin, J. Chem. Phys, **97**, 1954 (1992)

56. (a) A. I. Shushin, Chem. Phys. Lett. **118**, 197 (1985), (b) J. Chem. Phys. **95**, 3657 (1991).

57. A. I. Shushin, J. Chem. Phys, **113**, 4305 (2000); J. Chem. Phys. **110**, 12044 (1999).

58. N. Agmon, Phys. Rev. Lett. **67**, 1366 (1991); E. B. Krissinel, N. Agmon, J. Comp. Chem. **17**, 1085 (1996) *and references therein.*

59. E. R. Barthel, I. B. Martini, E. Keszei, and B. J. Schwartz, J. Chem. Phys. **118**, 5916 (2003)

60. K. H. Schmidt, P. Han, and D. M. Bartels, J. Phys. Chem. **99**, 10530 (1995).

61. K. H. Schmidt and D. M. Bartels, Chem. Phys. **190**, 145 (1995).

62. R. C. Rice, Diffusion-Limited Reactions (Elsevier, New York, 1985).

63. E. J. Hart and M. Anbar, The Hydrated Electron (John Wiley, New York, 1970).

64. A. J. Elliot, Can. J. Chem. **70**, 1658 (1992); H. A. Schwartz and H. J. Bielski, J. Phys. Chem. **90**, 1445 (1986); V. Nagarajan and R. W. Fessenden, J. Phys. Chem. **89**, 2330 (1985).

65. A. J. Elliot and F. C. Sopchyshyn, Int. J. Chem. Kinet. **16**, 1247 (1984); R. Devonshire and J. J. Weiss, J. Phys. Chem. **72**, 3815 (1968); D. Zehavi and J. Rabani, J. Phys. Chem. **76**, 312 (1972).

66. (a) Result described in a footnote in ref. 6. (b) At the higher end of the specified iodide concentration range, data are shown and discussed ref. 6. Unpublished data for 0.1 mM to 15 mM iodide are briefly discussed in the forthcoming paper by Chen et al. (ref. 10).

67. M. C. Sauer, Jr., R. Lian, R. A. Crowell, D. M. Bartels, and I. A. Shkrob*, to be submitted to J. Phys. Chem. A.*

68. R. Lian, R. A. Crowell, I. A. Shkrob, D. M. Bartels, D. A. Oulianov. and D. Gosztola, Chem. Phys. Lett., *submitted*; preprint available on http://www.arXiv.org/abs/physics/0401057.

69. C. Bressler, M. Saes, M. Chergui, D. Grolimund, R. Abela, and P. Pattison, J. Chem. Phys. **116**, 2955 (2002).

70. I. A. Shkrob, D. A. Oulianov, R. A. Crowell, and S. Pommeret, J. Appl. Phys., *submitted;* preprint available on http://www.arXiv.org/abs/physics/0401039 and http://www.arXiv.org/abs/physics/0401040.

71. e.g., P. M. Solomon, Nature **217**, 334 (1968).

72. K. J. Reed, A. H. Zimmerman, H. C. Andersen, and J. I. Brauman, J. Chem. Phys. **64**, 1368 (1976).

73. C. J. Smallwood, W. B. Bosma, R. E. Larsen, and B. J. Schwartz, J. Chem. Phys. **119**, 11263 (2003).

74. B. J. Schwartz, *private communication.*

75. P. Clifford, N. J. B. Green, and M. J. Pilling, J. Phys. Chem. **86**, 1318 and 1322 (1982); S. M. Pimblott, J. Phys. Chem. **95**, 6946 (1991).

76. T. Ichino and R. W. Fessenden, J. Phys. Chem. A. **107**, 9257 (2003); Y. Kajii and R. W. Fessenden, Res. Chem. Intermed. **25**, 567 (1999); A. S. Jeevarajan and R. W. Fessenden, J. Phys. Chem. **96**, 1520 (1992); R. W. Fessenden and N. C. Verma, J. Am. Chem. Soc. **98**, 243 (1976).

77. A. Bussandri and H. van Willigen, Chem. Phys. Lett. **344**, 49 (2001); J. Phys. Chem. A **106**, 1524 (2002); ); J. Phys. Chem. A **105**, 4669 (2001).

78. I. A. Shkrob, Chem. Phys. Lett. **264**, 417 (1997).

**Figure captions.**

**Fig. 1.**

(a) Transient absorption kinetics obtained in 200 nm pump (320 fs FWHM) - 800 nm (50 fs FWHM) probe experiments on 9.8, 56, and 98 mM aqueous solutions of KOH (150 μm jet; 200 nm photon fluence of $1.6 \times 10^{-2}$ J/cm$^2$). These kinetics are normalized at the maximum and given on logarithmic time scale. Only hydrated electron absorbs at 800 nm. The vertical bars represent 95% confidence limits for each data point. For the lowest OH$^-$ concentration, the maximum $\Delta OD$ signal was ca. $6 \times 10^{-3}$; the signal from 2-photon ionization of the solvent was ca. $3 \times 10^{-4}$. For the other two solutions, the maximum signals were $3 \times 10^{-2}$ and $5.5 \times 10^{-2}$, respectively. (b) Transient absorption resulting from 200 nm photoexcitation of 56 mM KOH. The probe wavelengths, *viz.* 650, 800, and 1000 nm, were selected from white light supercontinuum using 10 nm FWHM interference filters. These kinetic traces were normalized at 5 ps; the dashed line corresponds to zero absorption signal. The difference in each trace during the earliest phase is due to rapid blue shift in the electron spectrum that occurs in the course of electron thermalization (see section III.A).

**Fig 2.**

(a) Effect of the 200 nm pump power on the normalized 800 nm kinetics for hydrated electron in 98 mM aqueous KOH. Traces (i) and (ii) were obtained with 3.5 and 10 μJ excitation pulses, respectively. The maximum $\Delta OD$ signals were $1.5 \times 10^{-2}$ and $8.5 \times 10^{-2}$, respectively. No change in the kinetics as a function of pump power for 200 nm photon fluence $< 3 \times 10^{-2}$ J/cm$^2$ was observed. (b) The $t=10$ ps absorbance is plotted against the 200 nm photon fluence. This dependence is linear which suggests monophotonic excitation of OH$^-$ by the pump.

**Fig. 3.**

(a) Temperature dependencies of the 800 nm absorbance observed in 200 nm photoexcitation of 56 mM KOH (8 to 97 °C; see (b) for the long-term kinetics). The origin of the y-axis corresponds to zero absorption signal. The temperatures in °C are given in the color table to the right of the plot. To facilitate the comparison, the traces were normalized in the time interval indicated with a rectangle; at these delay times, the thermalization dynamics of the electron are complete (Fig. 1(b)). These dynamics manifest themselves at this probe wavelength as a fast decay phase observed in the first 4 ps. The amplitude of this decay decreases with the temperature because the absorption spectrum of a *thermalized* electron shifts to the red and the magnitude of the blue shift that occurs during thermalization becomes smaller. In (b), normalized 200 nm pump – 800 nm probe kinetics from 56 mM KOH as a function of the temperature are given. The same kinetics, temperatures, and color coding aplly as in (a). Note the logarithmic time scale. The vertical bars are 95% confidence limits for each data point; solid lines are the least-squares fits obtained using the equations given in section IV.A for $v \gg 1$. The fit parameters are plotted in Fig. 7. In (c), the ratio of $\Delta OD$ signals attained at 550 ps and (i) 5 ps and (ii) at the maximum are plotted as a function of the temperature. Both of these quantities (which are related to the survival probability of the geminate pair) increase linearly with the temperature, increasing by a factor of 3 over the 100 °C interval.

**Fig. 4.**

Normalized 200 nm pump (300 fs FWHM; 7 μJ; 170 μm) – 800 nm probe kinetics for 0.056, 0.99, 4.8, and 9.74 M KOH. The penetration depth of the 200 nm light was 78, 4.4, 0.9, and 0.45 μm, respectively. The concentrated alkaline solutions were prepared by adding KOH pellets to the 0.99 N volumetric standard. The densities, bulk viscosities, and maximum $\Delta OD$ signals for the 0.99, 4.8, and 9.74 M solutions are, respectively, 1.05 g/cm$^3$, 1.0 cP, and 41 mOD, 1.21 g/cm$^3$, 1.61 cP, and 32 mOD, and 1.4 g/cm$^3$, 3.4 cP, and 24 mOD. While there is a small change in the thermalization dynamics of the electron for $t<5$ ps, at later delay times, the kinetics do not change with the OH$^-$ concentration (within the 95% confidence bars shown in (a)). In (b), these bars were not shown.

**Fig. 5.**

A comparison between (i) 200 nm pump – 800 nm probe (100 mM NaOH) and (ii) 389 nm pump – 650 nm probe (2 M NaOH) kinetics obtained in, respectively, one- and two-photon excitation of OH$^-$. See section III.B for more detail. The solid lines are biexponential least-squares fits in which the fast and slow exponential components have the same time constants (12.4 and 83.2 ps, respectively) but different weights and constant offsets (64:16:20 [fast:slow:constant] and 43:21:36 for traces (i) and (ii), respectively). See Fig. 9(a) for more detailed analysis using the formulas given in section IV. (b) A comparison between the normalized kinetics obtained in biphotonic 389 nm (solid line) and 337 nm (open circles) photoexcitation of 2 and 1 M NaOH, respectively. See the text for more detail. The inset (c) shows a pump intensity dependence for the solvated electron signal recorded at 10 ps delay after the 389 nm pump pulse, using a 650 nm probe. The double logarithmic plot of $\Delta OD$ vs. the pump irradiance yields the photon order $n$=1.86.

**Fig. 6.**

(a) Least-squares fit (using eqs. (14) to (17)) to a typical 200 nm pump – 800 nm probe kinetics for (OH:$e_{aq}^-$) pairs in a dilute aqueous OH$^-$ solution at 25 °C (in this case, 56 mM KOH). The fit parameters are given in the text. In (b), the "best" potential $u(r)$ given by eq. (19) that yields these optimum fit parameters is plotted. The vertical dashed line indicates the Onsager radius $a$.

**Fig. 7.**

Analysis of the kinetic data given in Fig. 3(b) using Shushin's theory for diffusion controlled reactions in a potential well (section IV.A). Temperature dependencies for (a) the dimensionless parameter $\alpha$ (filled circles, to the left), the lifetime $W^{-1}$ of the hydrated electron in the potential well (empty squares, to the right) and (b) the escape probability $p_d$ from the well are given. The vertical bars are 95% confidence limits; for $p_d$ the error is smaller than the symbols.

**Fig. 8.**

Data of Fig. 7 replotted in a different parameter space, as explained in section IV.A. (a) Temperature dependencies for the Onsager radius $a$ (filled circles; the vertical bars give 95% confidence limits) and the effective radius $R_{eff}$ of rxn. (2) in the bulk (empty squares) calculated using eq. (18). (b) *To the right:* Temperature plots *(solid lines)* for $W_r$ (filled circles) and $W_d$ (filled squares). The estimated activation energies are given in the text. *To the left:* $W_d/D$ (empty squares) as a function of temperature. The diffusion coefficient $D_e$ for hydrated electron given in Fig. 7S (ref. [60]) has been assumed for $D$.

**Fig. 9.**

(a) Data of Fig. 5 reanalyzed using the model given in section IV.A. The time is given in the units of $W^{-1}$, i.e., $\tau = Wt$. The optimum parameter set from Fig. 6(a)): $p_d$=0.173, $\alpha$=0.381, and $W^{-1}$=12.75 ps was used; the only adjustable parameter was $\upsilon = a/b$, the ratio of the Onsager radius $a$ and the distribution width $b$ (eq. 22). For (i) $\upsilon \gg 1$; for (ii), $\upsilon$=3. The time-domain kinetics were obtained by numerical Laplace transform of $\Omega(s)$ obtained using eqs. (6), (9), (10), and (23). (b) The survival probability $\Omega(\tau)$ as a function of $\upsilon$ was calculated using the same model parameters as in (a). The dimensionless time $\tau$ is given on the logarithmic scale. Parameters $\upsilon$ are indicated in the plot next to the traces. For $\upsilon$<1, these kinetics look much like those given by eq. (8).

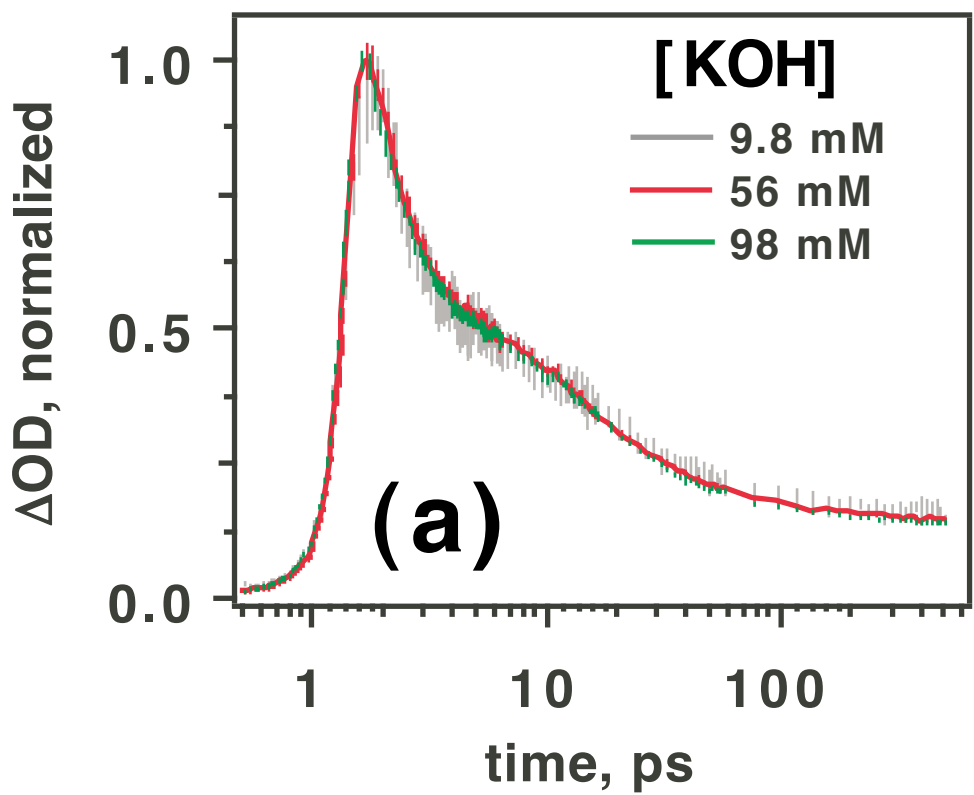
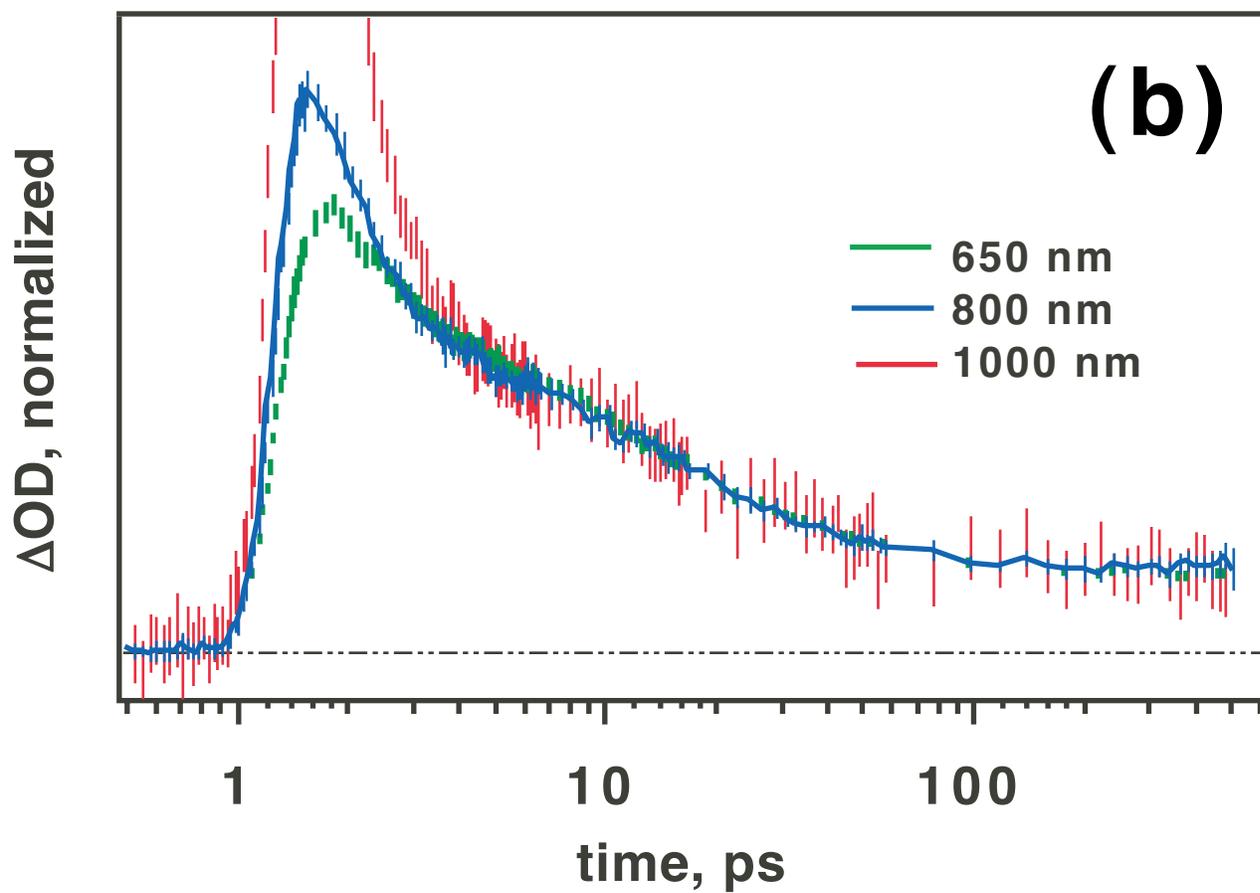

Crowell et al; Fig. 1

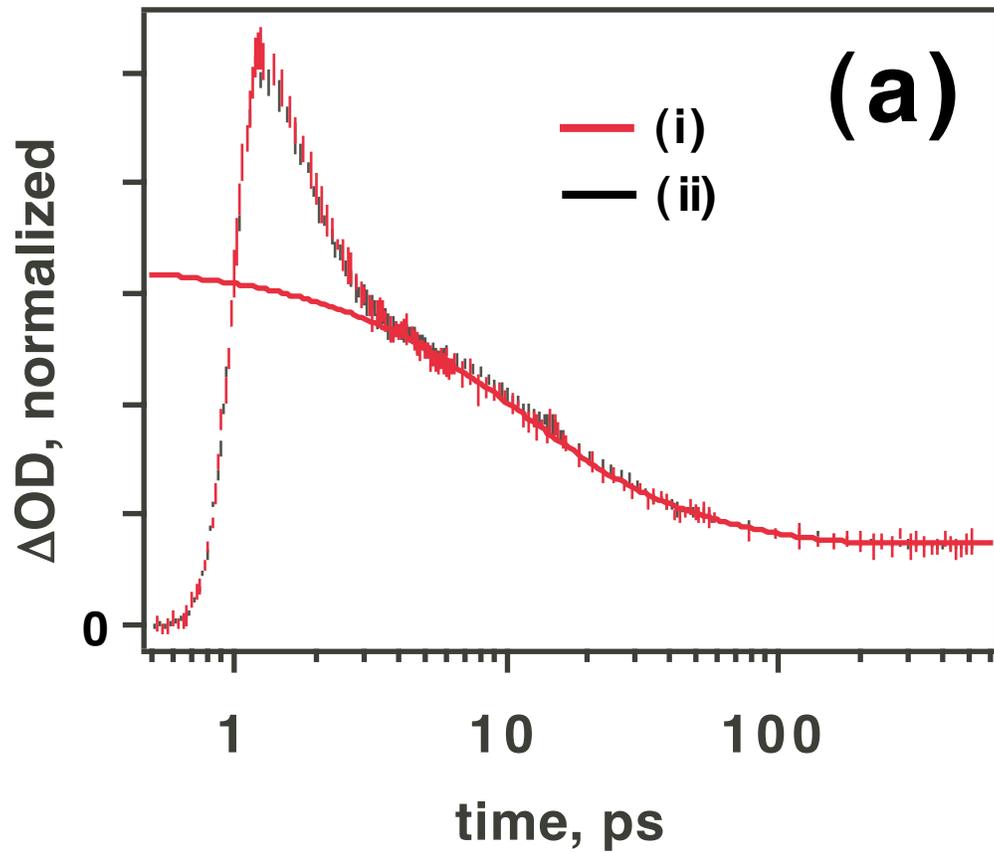
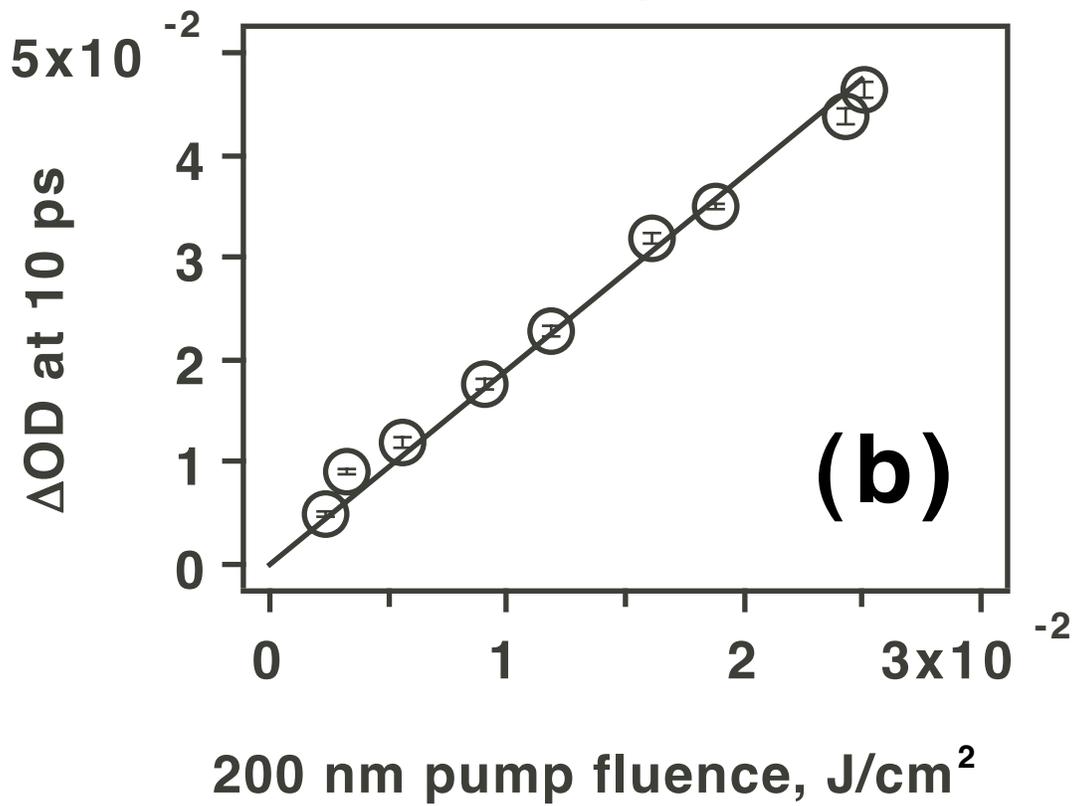

Crowell et al; Fig. 2

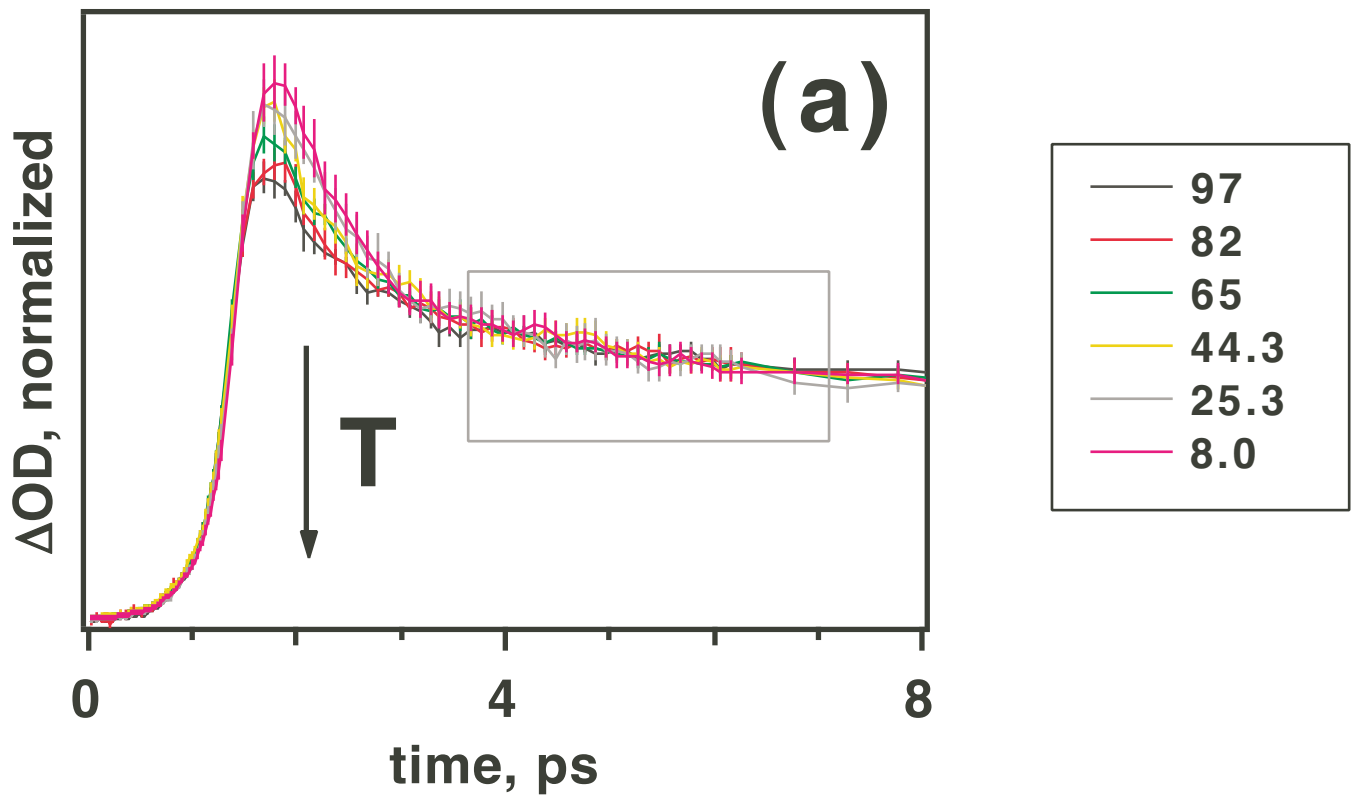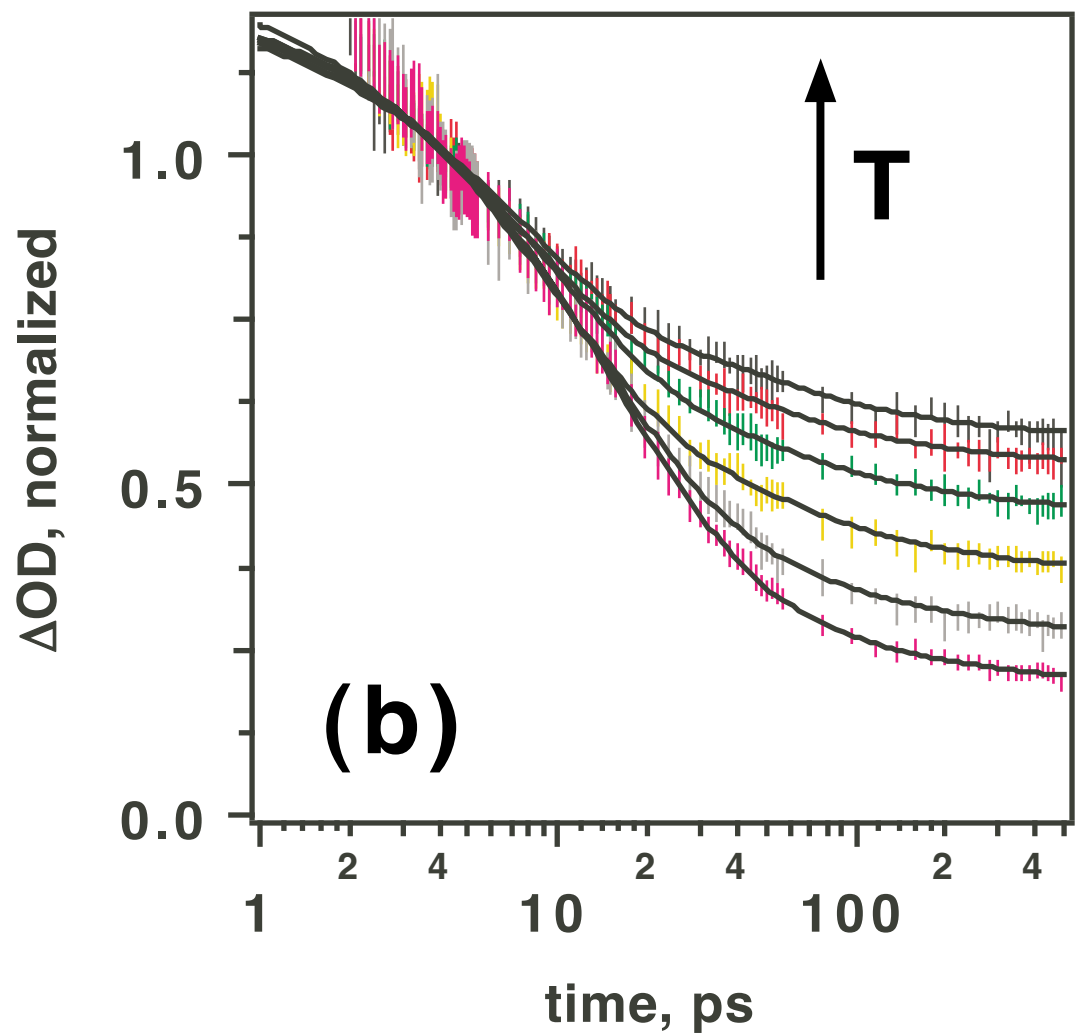

Crowell et al; Fig. 3a,b

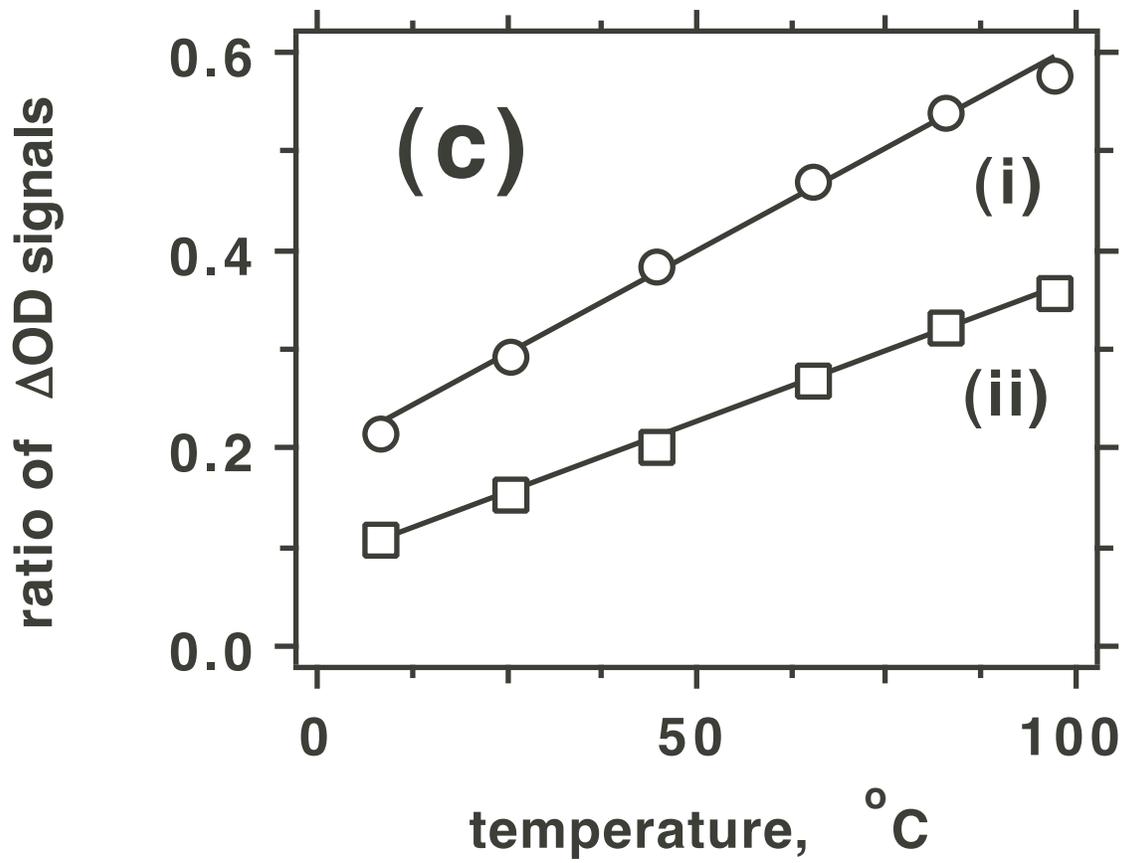

Crowell et al; Fig. 3c

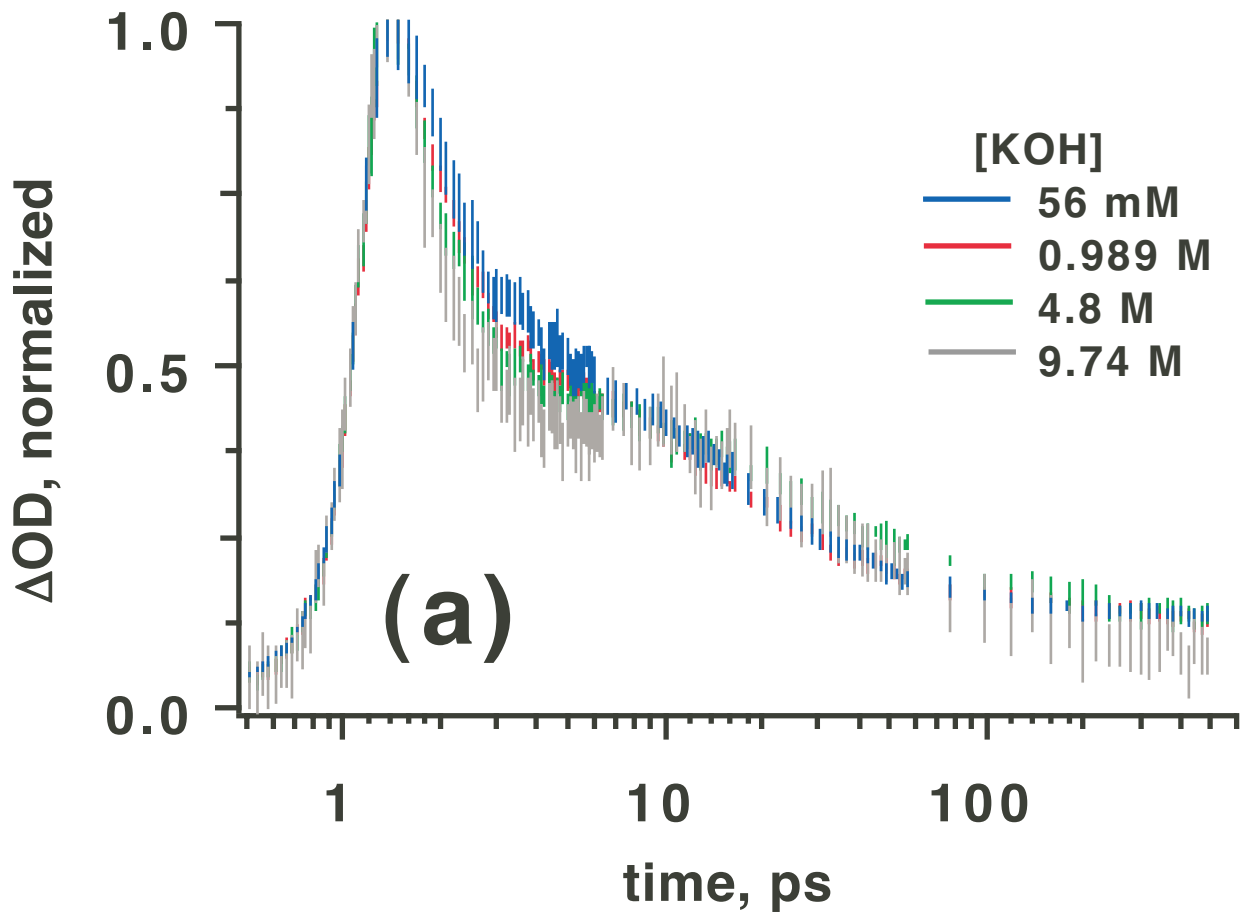
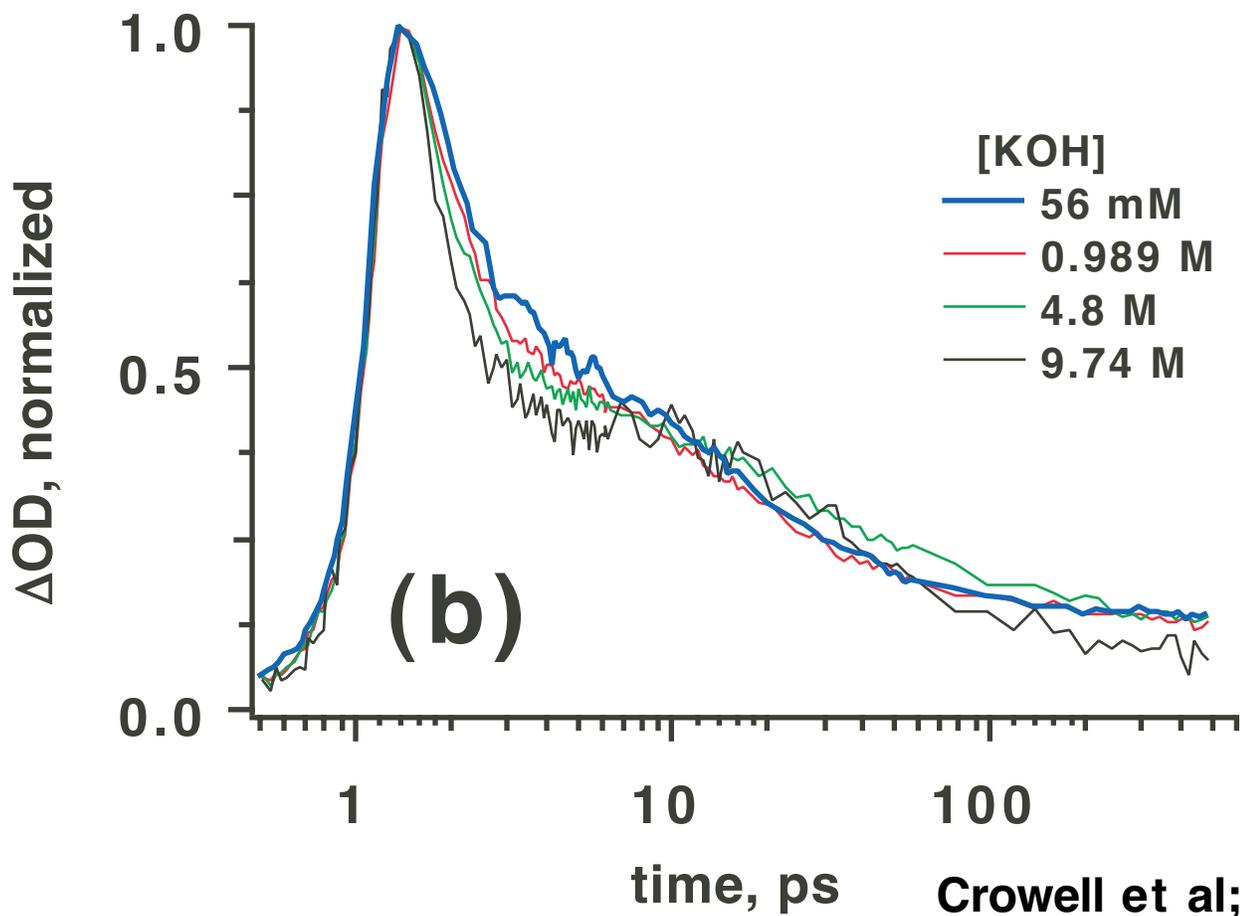

Crowell et al; Fig. 4

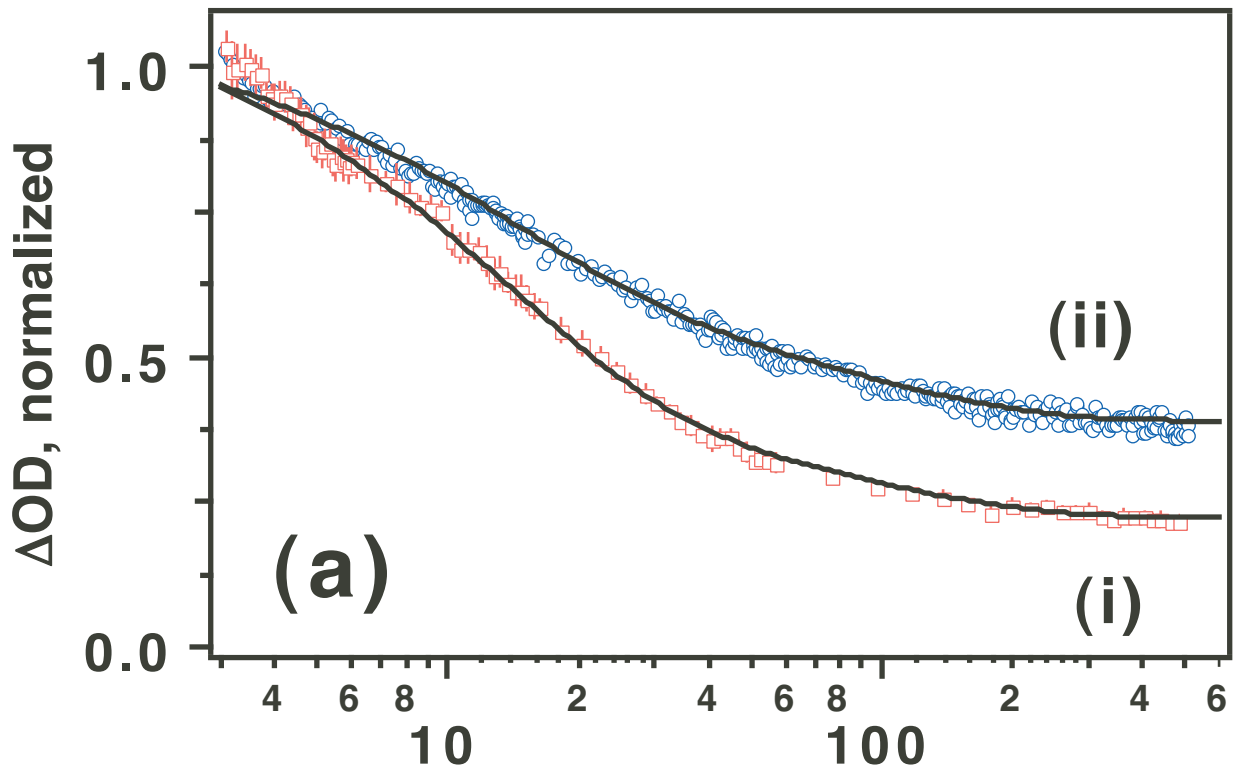
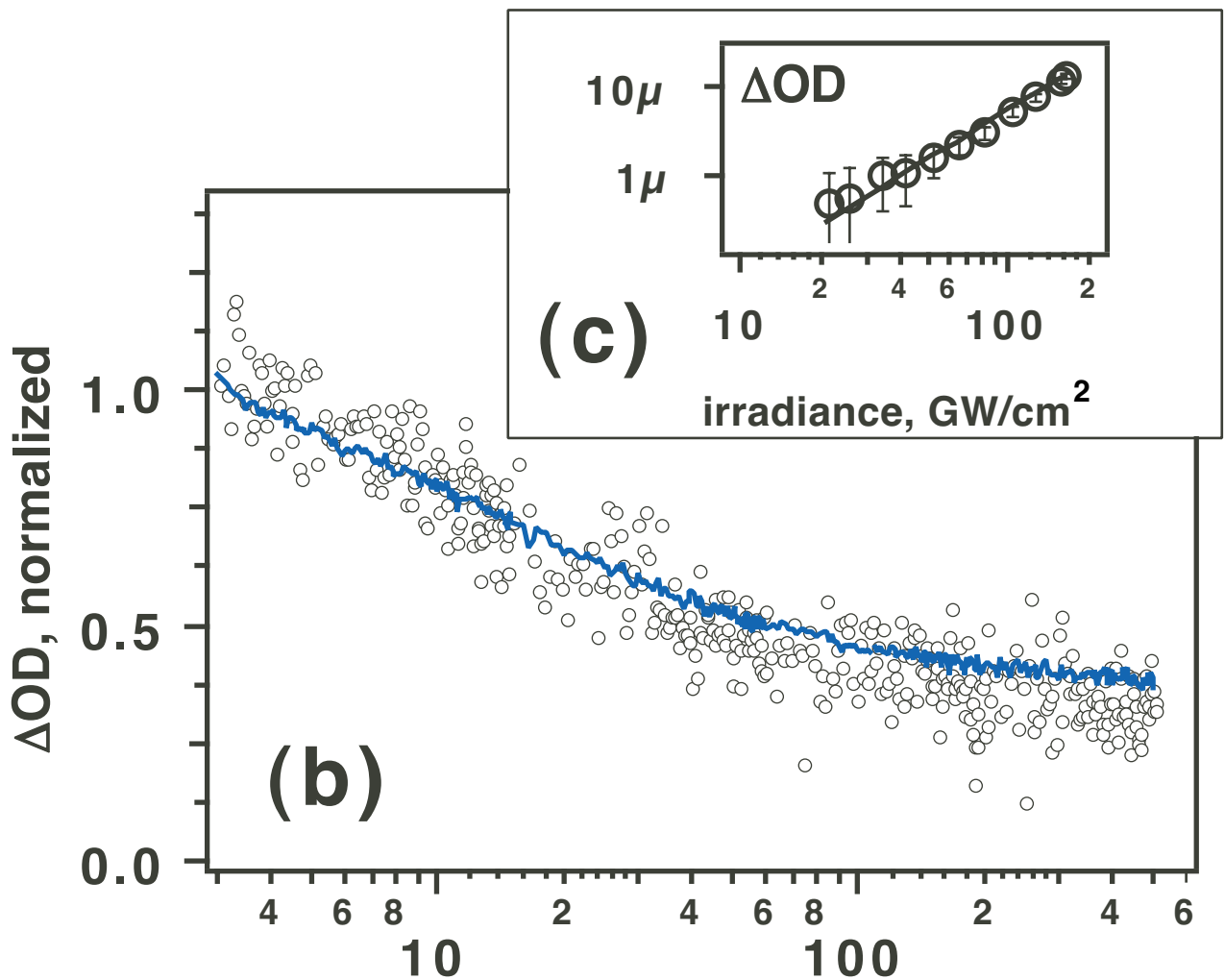

Crowell et al; Fig. 5

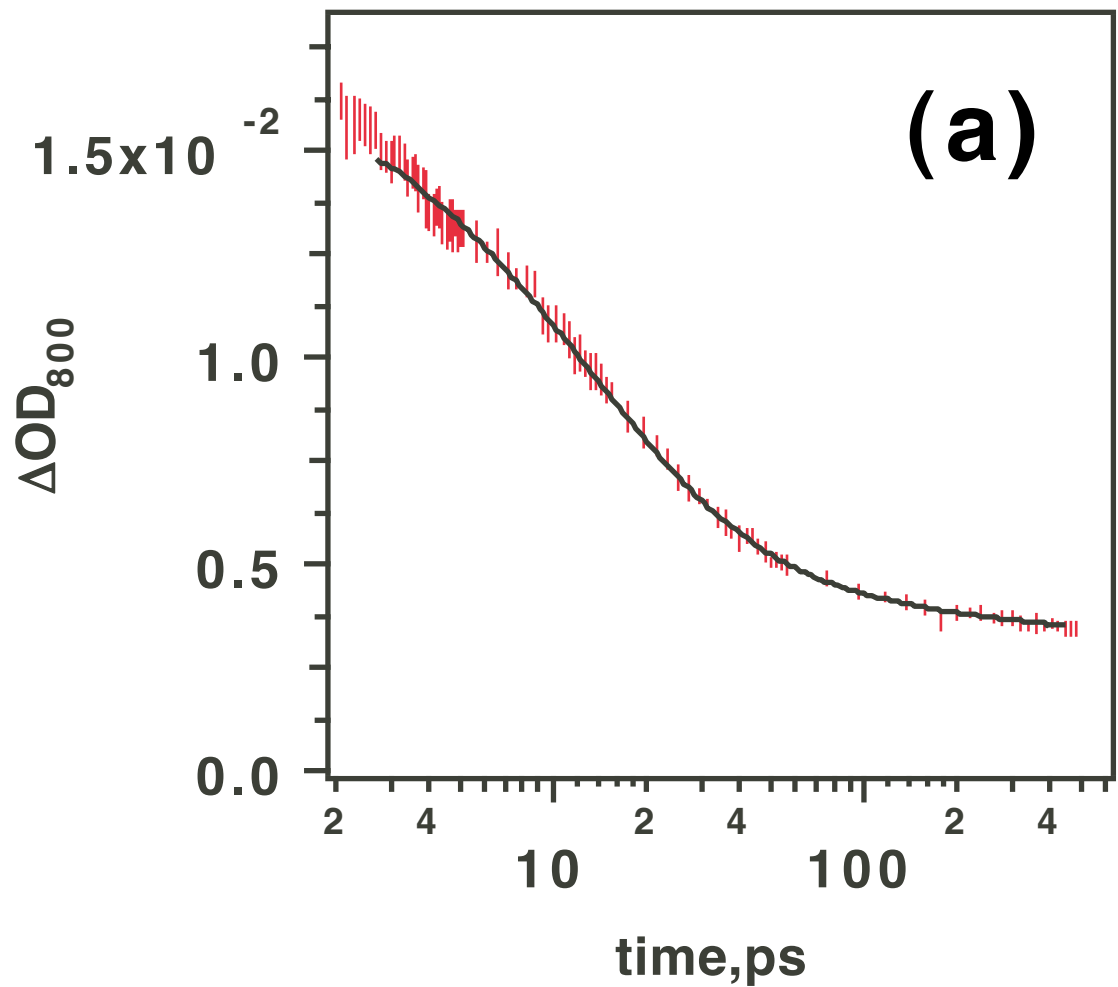

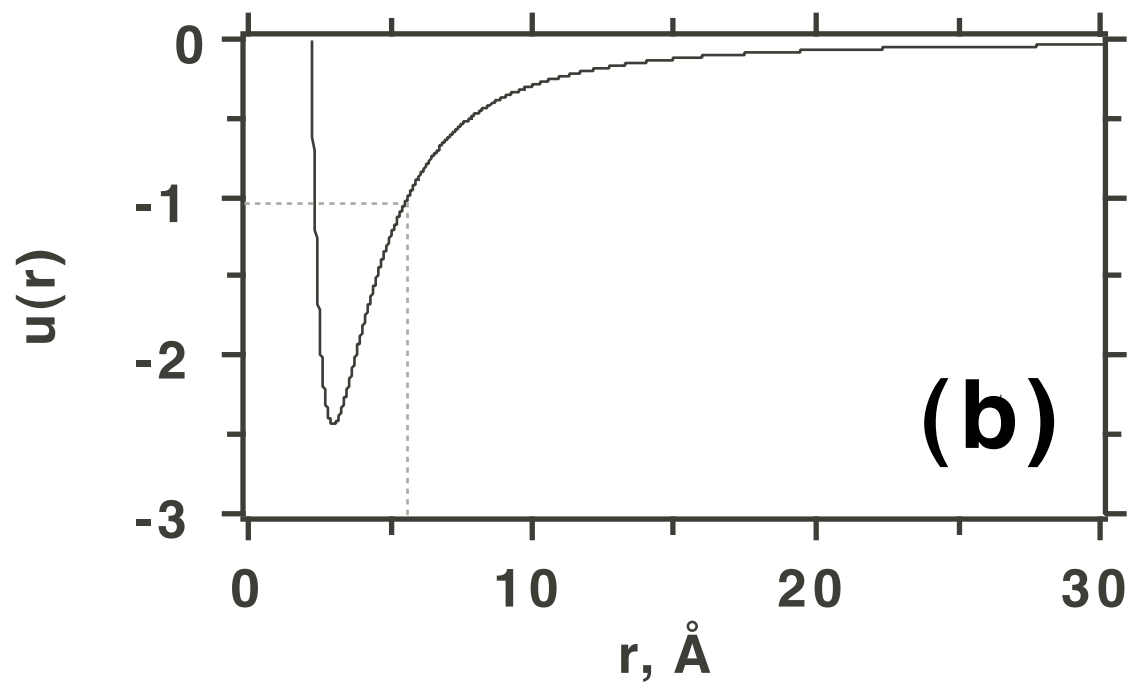

Crowell et al; Fig. 6

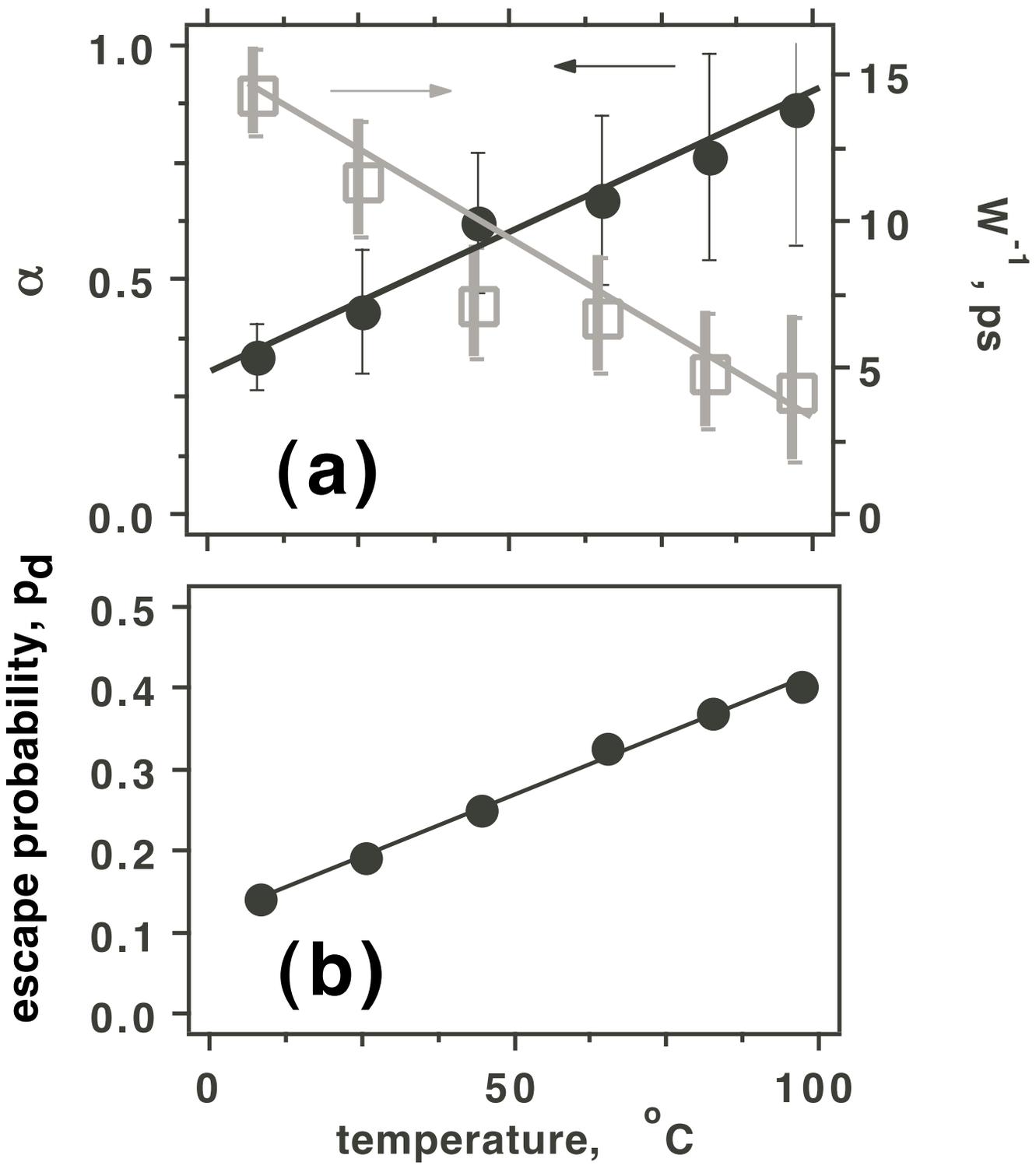

Crowell et al; Fig. 7

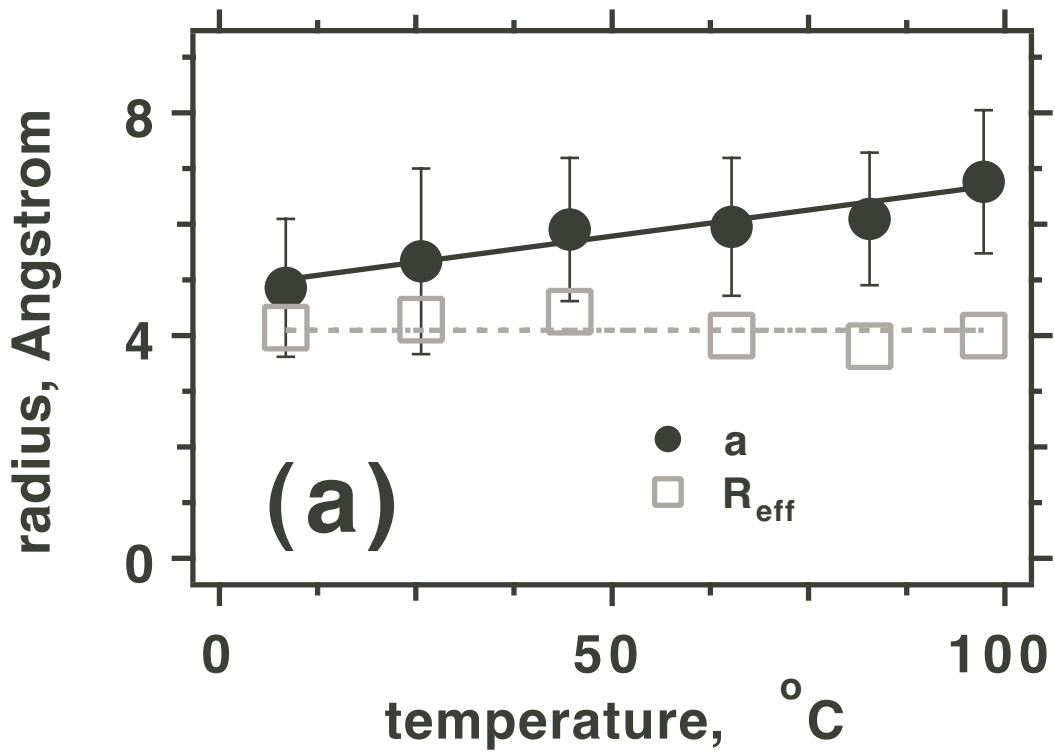
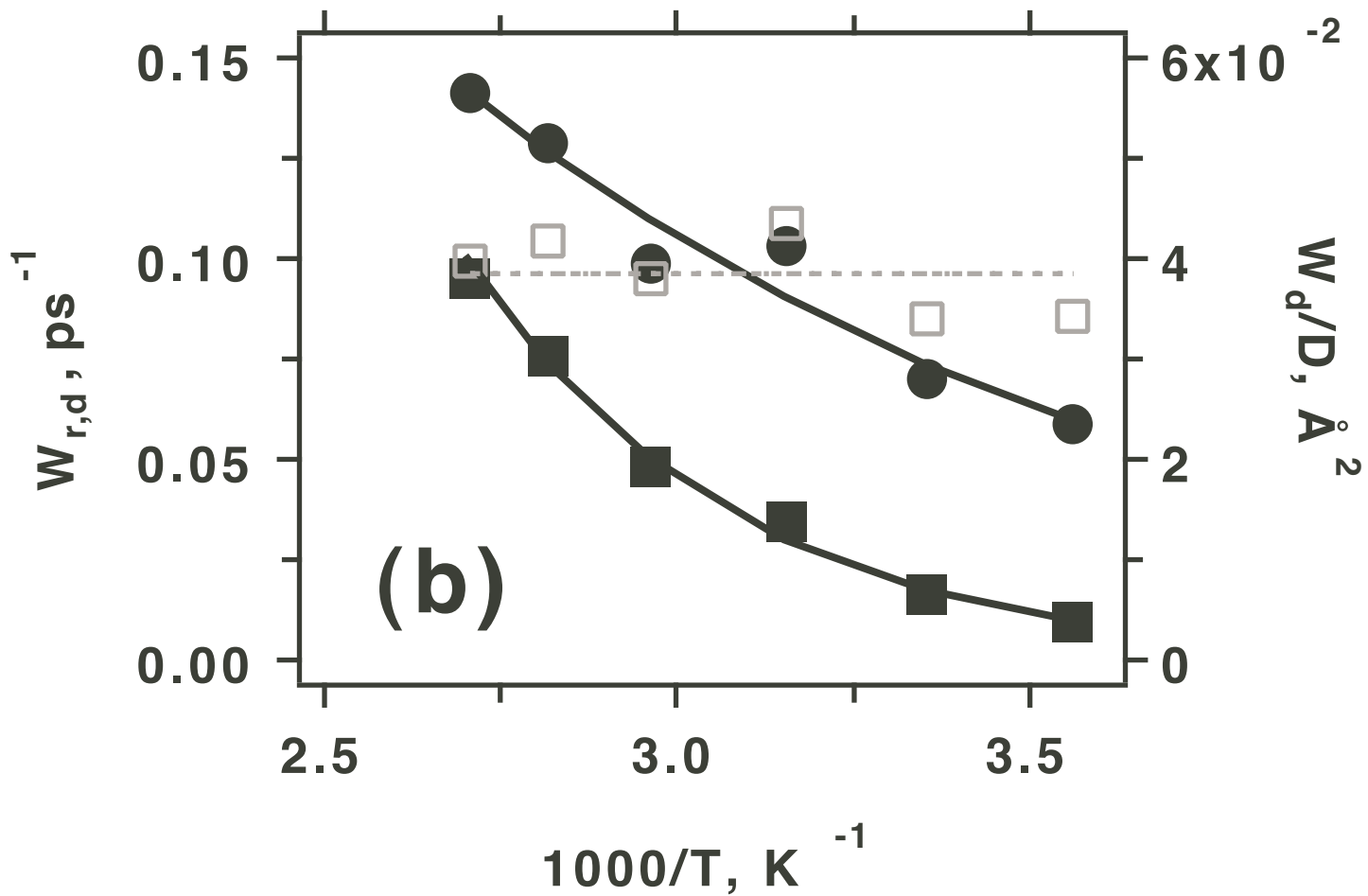

Crowell et al; Fig. 8

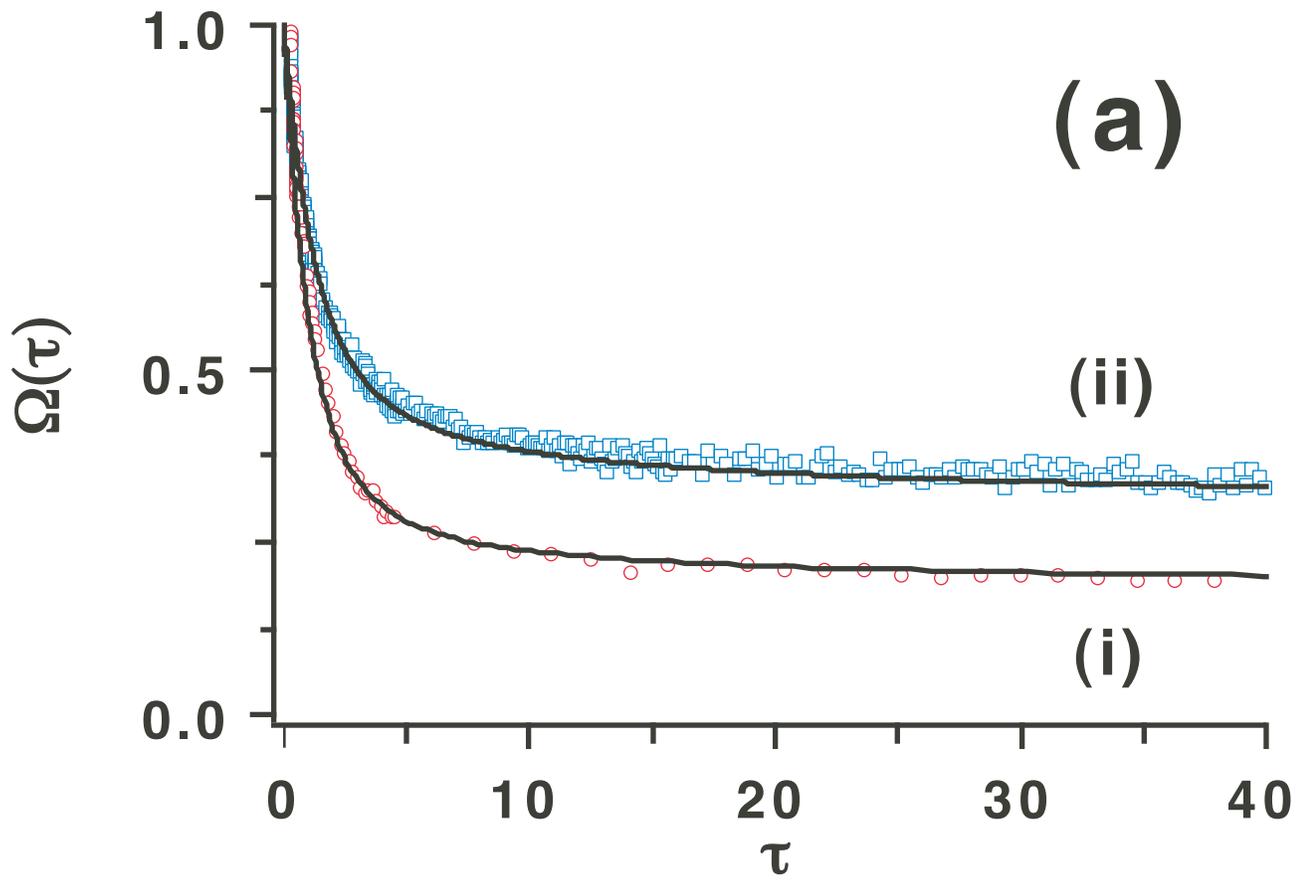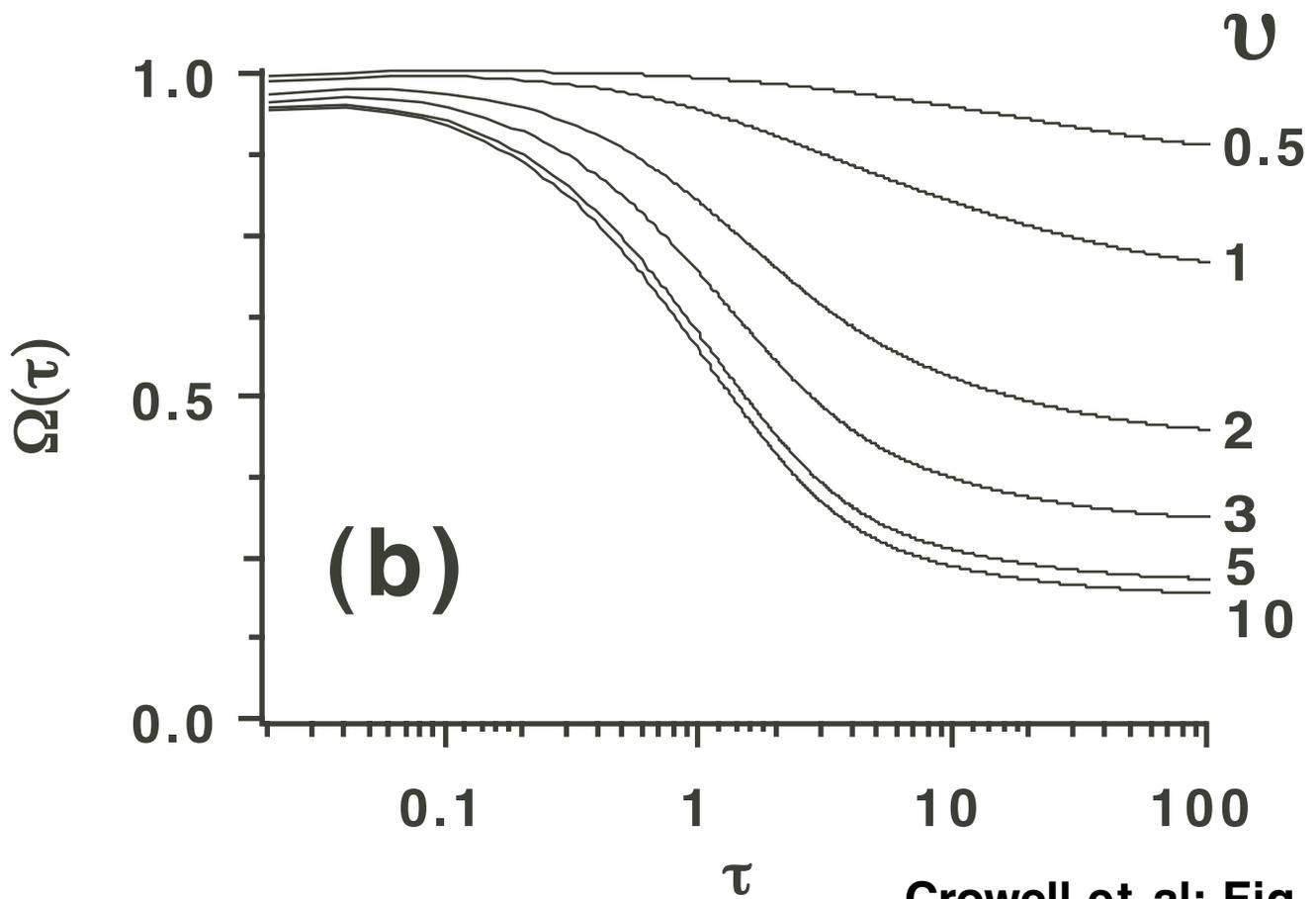

Crowell et al; Fig. 9

# Supporting Information.

## Appendix. Biphotonic excitation: the photophysics.

In this Appendix, we consider the photophysics of 400 nm excitation of aqueous hydroxide (Figs. 2S, 3S, and 4S). It is known that water itself can be photoionized by simultaneous absorption of three 400 nm photons. A 3-photon absorption coefficient $\beta_3$ of 900 cm$^3$/TW$^2$ has been reported for neat H$_2$O [50]. This estimate was obtained from transmission data on picosecond pulses passing through a 1 cm cell. Our recent study on electron yield and pump transmission for femtosecond 400 nm pulses passing through a 70 μm jet gives a somewhat lower estimate of 270 cm$^3$/TW$^2$ *(Crowell et al., in preparation)*. That study suggested that just below the onset of the dielectric breakdown of water at (5-10) TW/cm$^2$, the 3-photon regime is succeeded by a different regime (which was called "3+1" regime). The salient feature of the latter regime is that once the pump power exceeds a certain threshold level (0.3-1 TW/cm$^2$), the electron yield increases *linearly* with the incident laser power (e.g., Fig. 4S(a)), and the line does not extrapolate to zero yield for low irradiance. Simultaneously, the electron spectrum observed for $t>5$ ps shifts to the red and thermalization dynamics observed to the red of 800 nm become progressively slower; both of these changes correlate with the laser power. As shown elsewhere, all of these spectral and kinetic changes indicate rapid heat elimination by photoexcited water molecules that do not undergo ionization. Such a situation is possible due to extremely nonuniform deposition of photoexcitation energy across the sample, rapid heat transfer in water, and relatively low 3-photon quantum yield (ca. 0.42) for 400 nm photoionization of water.

Provided that the 2-photon absorbance by OH$^-$ is small (as shown below), unless the concentration of hydroxide is very high, the 3-photon excitation of the solvent always competes with the 2-photon excitation of the solute.

Transient absorbance experiments at high 400 nm irradiance were carried out at Argonne. Pump irradiances to 2 TW/cm$^2$ were obtained by doubling in frequency the Ti:sapphire fundamental. Fig. 2S shows typical power dependencies obtained for $e_{aq}^-$ yield in KOH solution, for pump radiance below the "3+1" threshold; Fig. 4S shows the same dependence over a larger range of pump irradiances, for [OH$^-$]=1 M. Whereas in neat water, the 800 nm absorbance increases as power $n=2.96$ of the pump irradiance (Fig. 2S(a)), this exponent decreases to $n=2.37$ or 2.56 in the 0.99 M (Figs. 2S(b) and 4S(a)) solution and 2.14 in the 4.6 M solution (Fig. 2S(a)). At any given pump power, the 800 nm absorbance increases linearly with [OH$^-$] (inset in Fig. 2S(b)), although the slope decreases markedly with the pump power.

These trends can be understood assuming that there is a competition between 2-photon absorption of the 400 nm photons by OH$^-$ and 3-photon absorption of these photons by water. To estimate the 2-photon absorption coefficient $\beta_2$[OH$^-$] for OH$^-$ in 2 x 400 nm photoexcitation of hydroxide, we assumed that the probe beam is infinitely narrow (since the ratio of the probe and pump beam radii was < 0.2) The absorption of the time-dependent irradiance $J(x,t)$ of the pump pulse is given by

$$dJ/dx = -\beta_2[HO^-]\,J^2 - \beta_3\,J^3 \tag{A1}$$

where $\beta_3$ is the three-photon absorption coefficient for water that was determined in a separate experiment. Eq. (A1) was integrated numerically and then the sample-average rate $d[e_{aq}^-]/dt$ of the electron formation was determined from

$$d[e_{aq}^-]/dt = L^{-1}\int_0^L dx\; J^2\left\{ \tfrac{1}{2}\phi_2\beta_2[OH^-] + \tfrac{1}{3}\phi_3\beta_3\,J \right\} \tag{A2}$$

where $\phi_2$ and $\phi_3$ are the corresponding QY's. The last step of the computation is to integrate eq. (A2) over time, assuming a Gaussian time profile for the pump pulse (the final result depends only on the pulse shape; it does not depend on the actual pulse width). Under our assumptions, the optical density $\Delta OD$ due to the electron absorbance at the probe wavelength is given by

$$\Delta OD = \varepsilon_{probe}\,L\,\int_{-\infty}^{+\infty} dt\; dE/dt. \tag{A3}$$

Numerical simulations shown in Fig. 3S indicate that for moderate (peak) pump irradiances $J_{pump}$, $\Delta OD \propto J_{pump}^n$, where $n$ is an exponential factor between 2 (for high concentration of hydroxide, when the first term in the integral eq. (A2) is large) and 3 (for dilute solutions). For sufficiently high pump irradiance ($> 1$ TW/cm$^2$), the second term in eq. (A2) always dominates. To estimate $\beta_2$, we let $\beta_3 = 270$ cm$^2$/TW$^3$ and $\phi_3 = 0.42$, assumed that $\phi_2 = 1$, and matched $\beta_2$ to obtain (i) the best fit of $\Delta OD$ for a fixed pump radiance as a function of [OH$^-$] and (ii) the best fit of factor $n$ as a function of [OH$^-$]. This procedure gave a self-consistent estimate of 3.8 cm TW$^{-1}$ M$^{-1}$ (see Fig. 3S).

**Supplement Figure Captions.**

Fig. 1S.

Normalized difference traces $\Delta OD(t)-\Delta OD(t=550\ ps)$ from Fig. 3(b) as a function of temperature (indicated in the color table). The traces were normalized by the $t>50$ ps integral. Dashed horizontal line corresponds to zero.

Fig. 2S.

Double logarithmic plot of a transient 800 nm absorbance at $t=20$ ps vs. 400 nm pump power (given in arbitrary units; one unit corresponds to ca. 0.2 TW/cm$^2$) for (a) neat water (circles) and 4.6 M KOH (squares) and (b) 0.989 M (squares) and 5 M KOH (circles). The exponential factors $n$ are given in the plot. These factors depended on the experimental conditions (e.g., jet thickness, irradiance range, and beam spot matching) and were not perfectly reproducible from one series to another. The general trend is that the photon order $n$ systematically decreases from 3 to 2 with the increase in [KOH]. The inset in (b) shows $\Delta OD(t=20\,ps)$ for 0.13 TW/cm$^2$ pump as a function of [KOH]. The concentration plots are linear for other pump powers too.

Fig. 3S.

A simulation of the power dependencies shown in Fig. 2S using numerical solutions to eqs. (A1) and (A2) in the Appendix. The following parameters were assumed: optical path of 150 µm, molar extinction coefficient for the electron of 17,670 M$^{-1}$ cm$^{-1}$, $\beta_3$=270 cm$^2$/TW$^3$ for neat water, $\phi_3$ =0.42 (see Appendix) and $\phi_2$ =1. The absorption coefficient $\beta_2$[OH$^-$] was varied between 0 (trace (i)) and 50 cm/TW and the power plots were analyzed in the same way as the data in Fig. 2S. The value $\beta_2$=3.8 cm/TW gave the best match to the concentration dependence shown in Fig. 2S(b) and the observed dependence of $n$ vs. [KOH] (given in the color table to the right of the plot). For reference, the power dependence for a hypothetical photosystem in which all of the 400 nm light is absorbed by 1 M OH$^-$ is given in trace (ii).

Fig. 4S.

(a) Power dependence for 800 nm absorbance in 400 nm excitation of 0.989 M KOH. The signal is obtained at $t=7$ ps at the end of the thermalization dynamics of the electron. Note the double logarithmic scale. The origin of the y-axis corresponds to zero photoinduced absorbance. At the higher end of the available irradiance range (> 1 TW/cm$^2$), the photon order changes from 2.37 to 1.04. Over the same irradiance range, the photon order for neat water also changes from 3 to 1, indicating the onset of the "3+1" regime (see Appendix). (b) Normalized 800 nm kinetics obtained for 400 nm excitation of the same solution with (i) 3, (ii) 15.6, and (iii) 85 µJ pump pulses. At the highest power, the geminate kinetics are very flat; the gentle bend is due to cross recombination. At low and intermediate pump power, the kinetics are a weighted sum of the 2-photon kinetics shown in Fig. 5(a), trace (ii) and 3-photon kinetics due to the water ionization. In a thin jet, the 2-photon kinetics can be observed only when the absorption signal is low, < 10 µOD; otherwise, 3-photon ionization always intrudes. Fig. 5 show signals in this regime collected at higher repetition rate in a longer path length cell.

Fig. 5S.

$M(\alpha,\tau)$ given by eq. (17) vs. dimensionless time $\tau = Wt$ as a function of parameter $\alpha$. *From top to bottom: $\alpha$=0 to 1 in the increments of 0.1.*

Fig. 6S.

"Temperature dependencies" for parameters $D/W$, $\alpha$, and $p_d$ calculated using the potential at $T_o$=298 K which is shown in Fig. 6(b) (see section IV.A for the parameterization). The parameters shown in the temperature plot were obtained by scaling the potential $U(r)$ by a factor of $T_o/T$, as described in the text.

Fig. 7S.

Diffusion constant $D_e$ of hydrated electron in neat water plotted as a function of temperature. The empirical formula used to calculate these data (for $D_e$ in cm$^2$/s and the temperature $T$ in $^o$C) is given below (taken from ref. [60]).

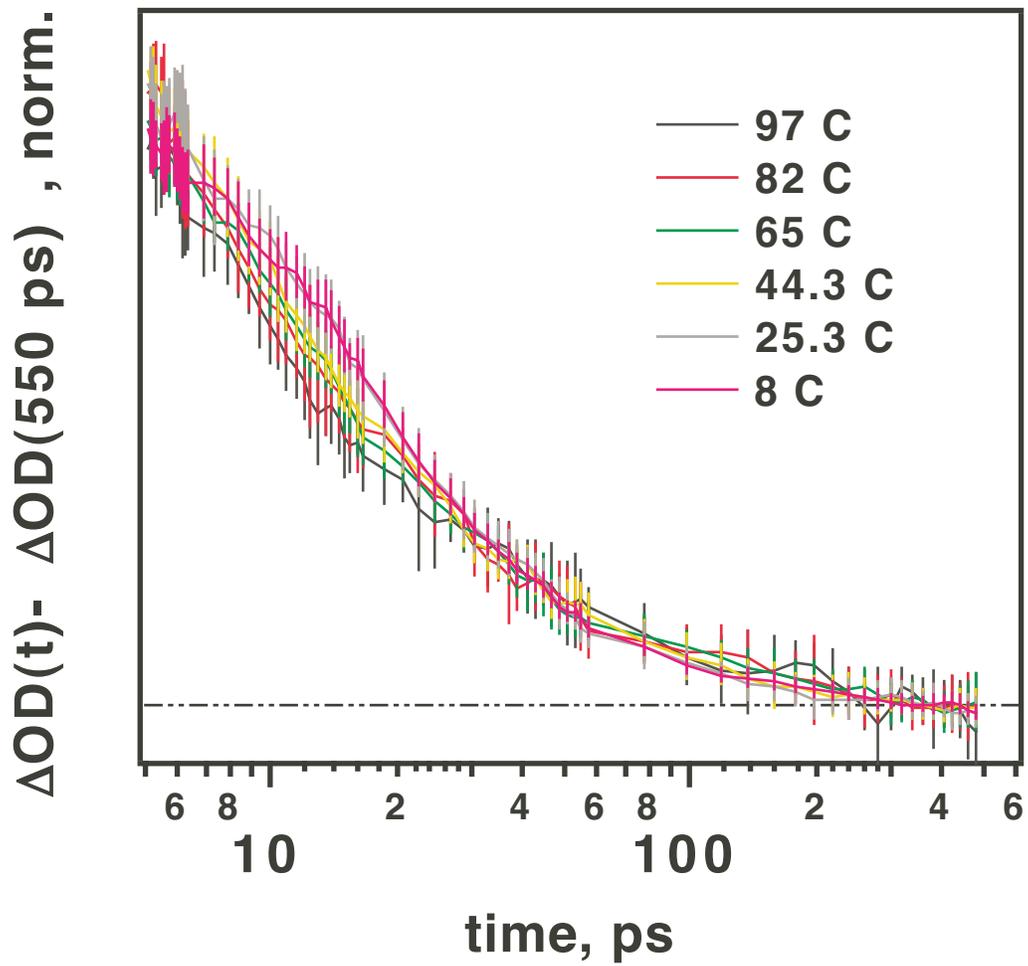

Crowell et al; Fig. 1S

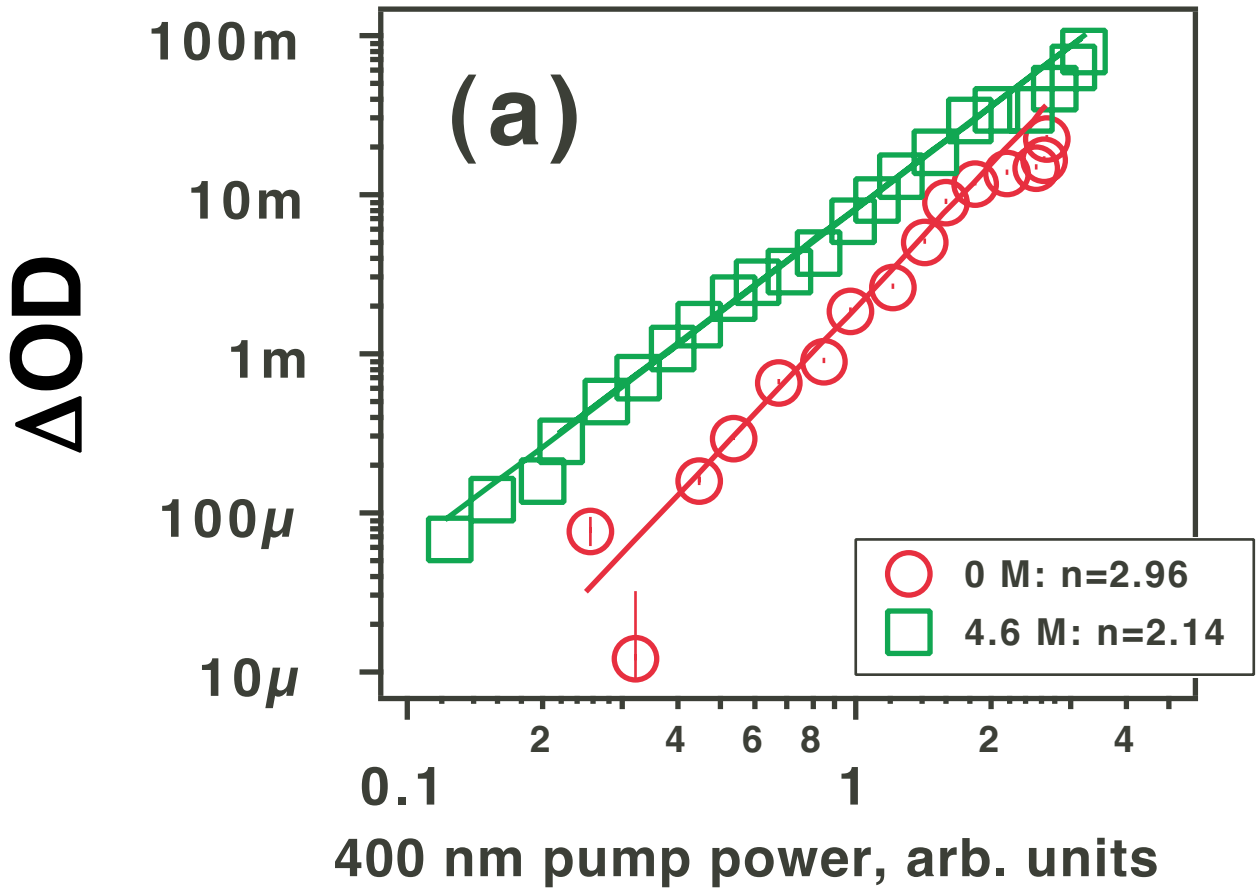
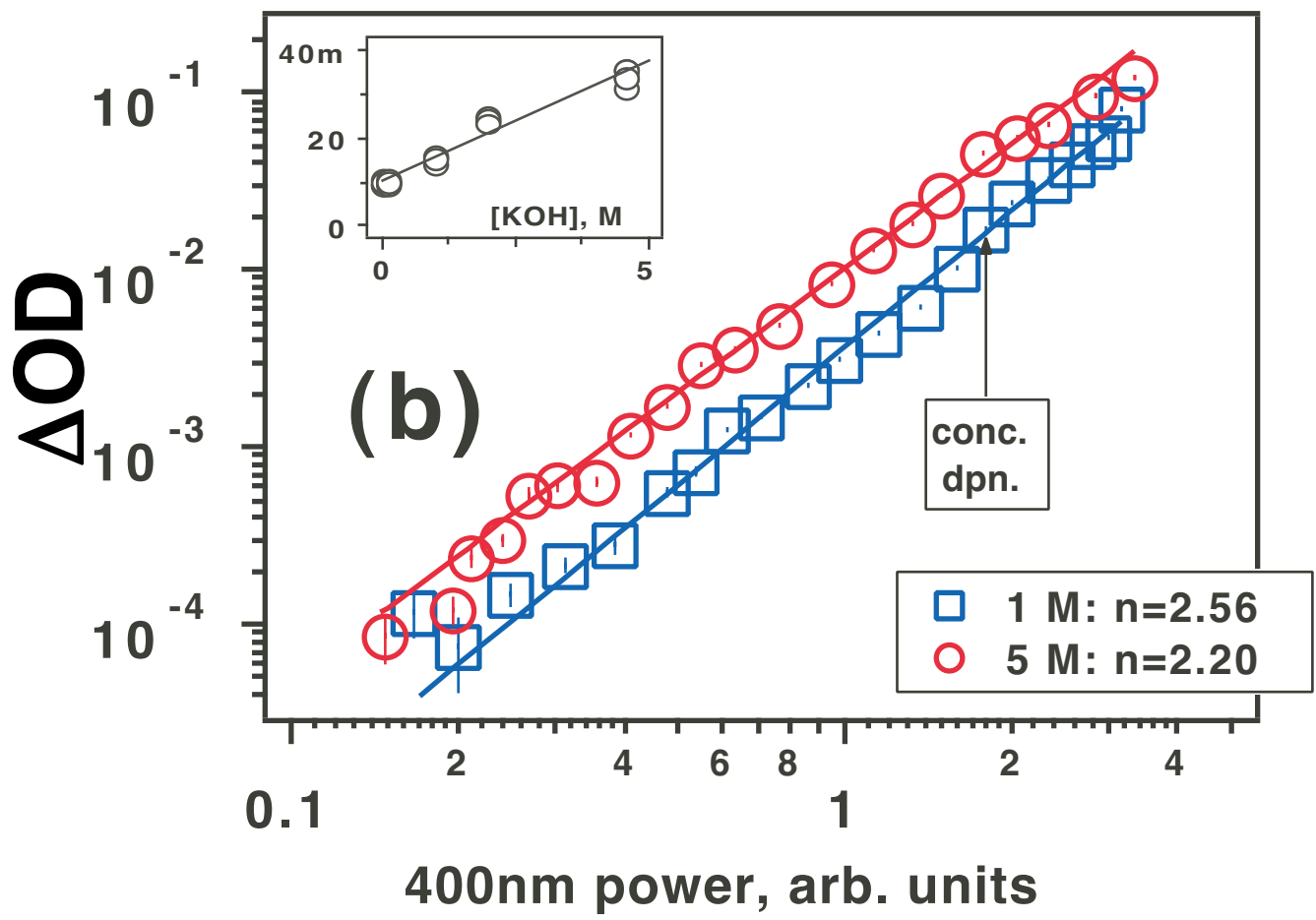

Crowell et al; Fig. 2S

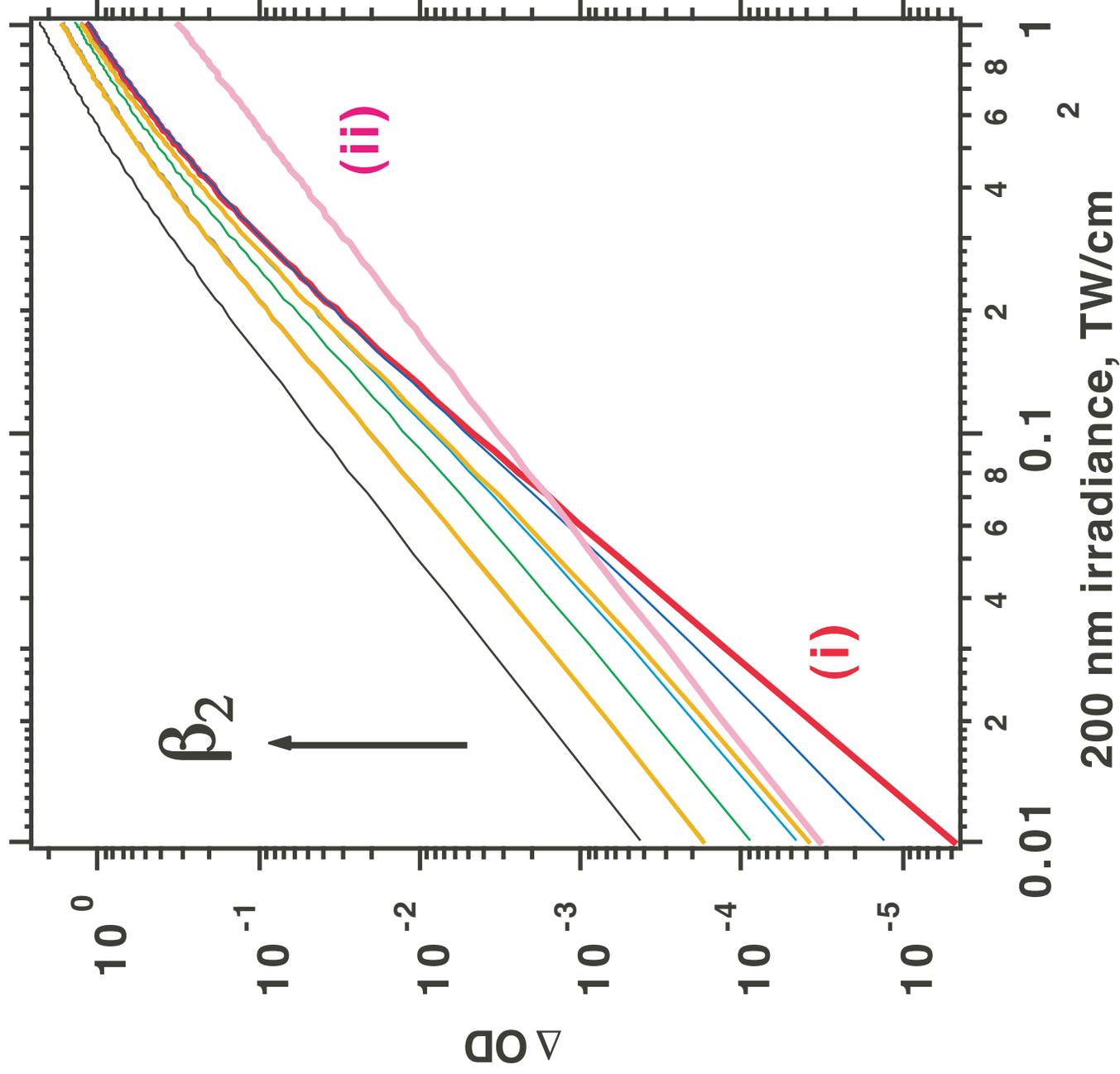

Crowell et al; Fig. 3S

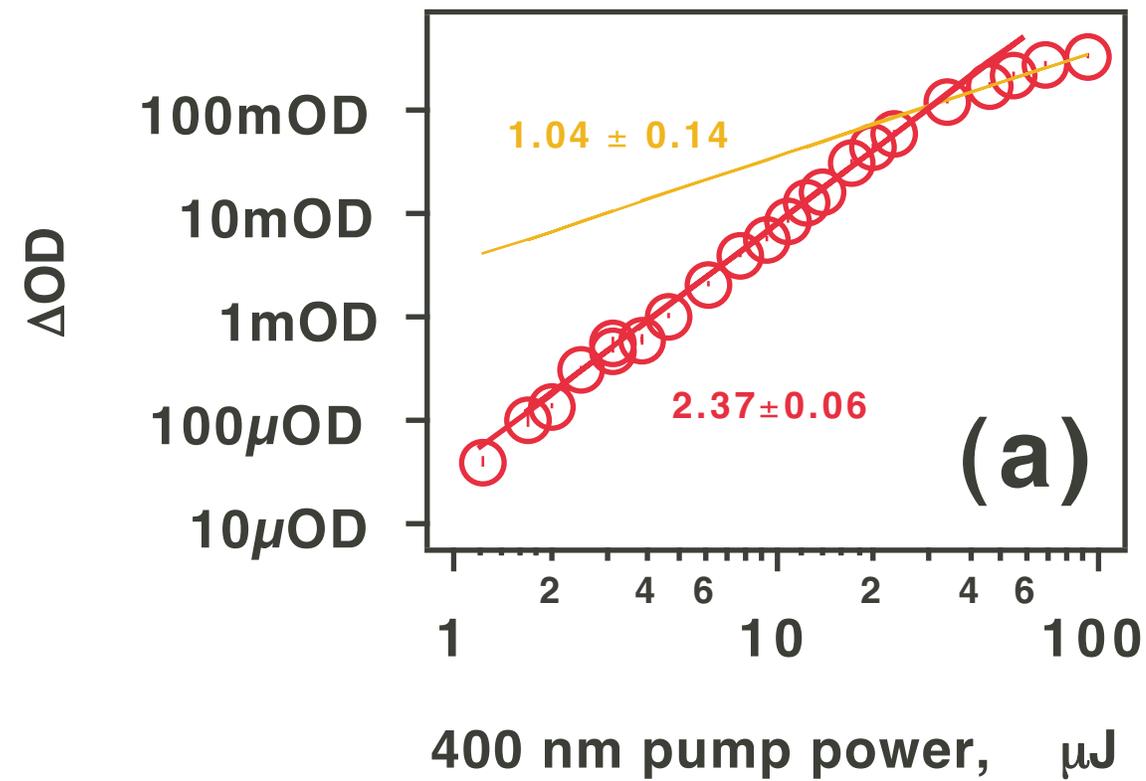
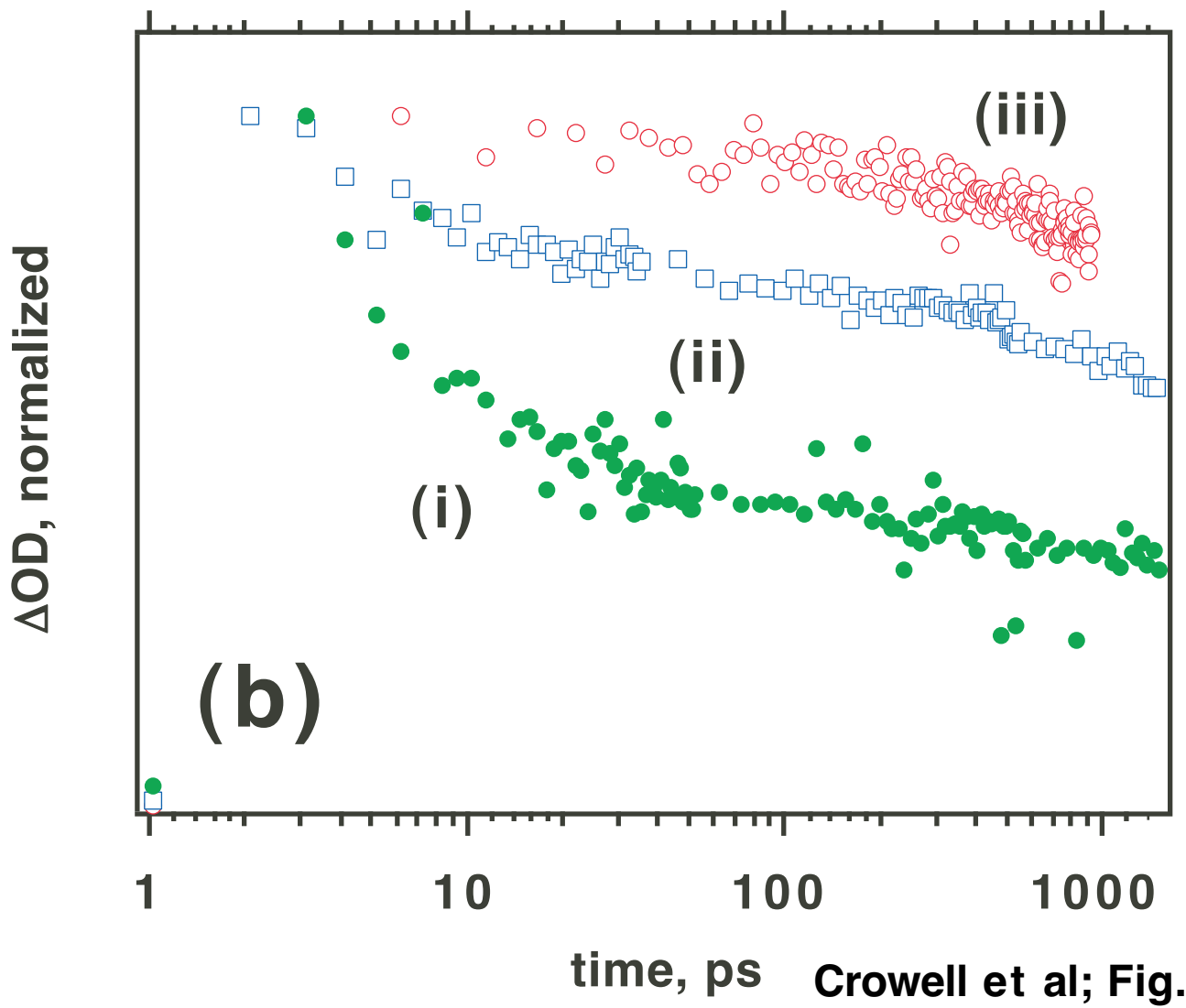

Crowell et al; Fig. 4S

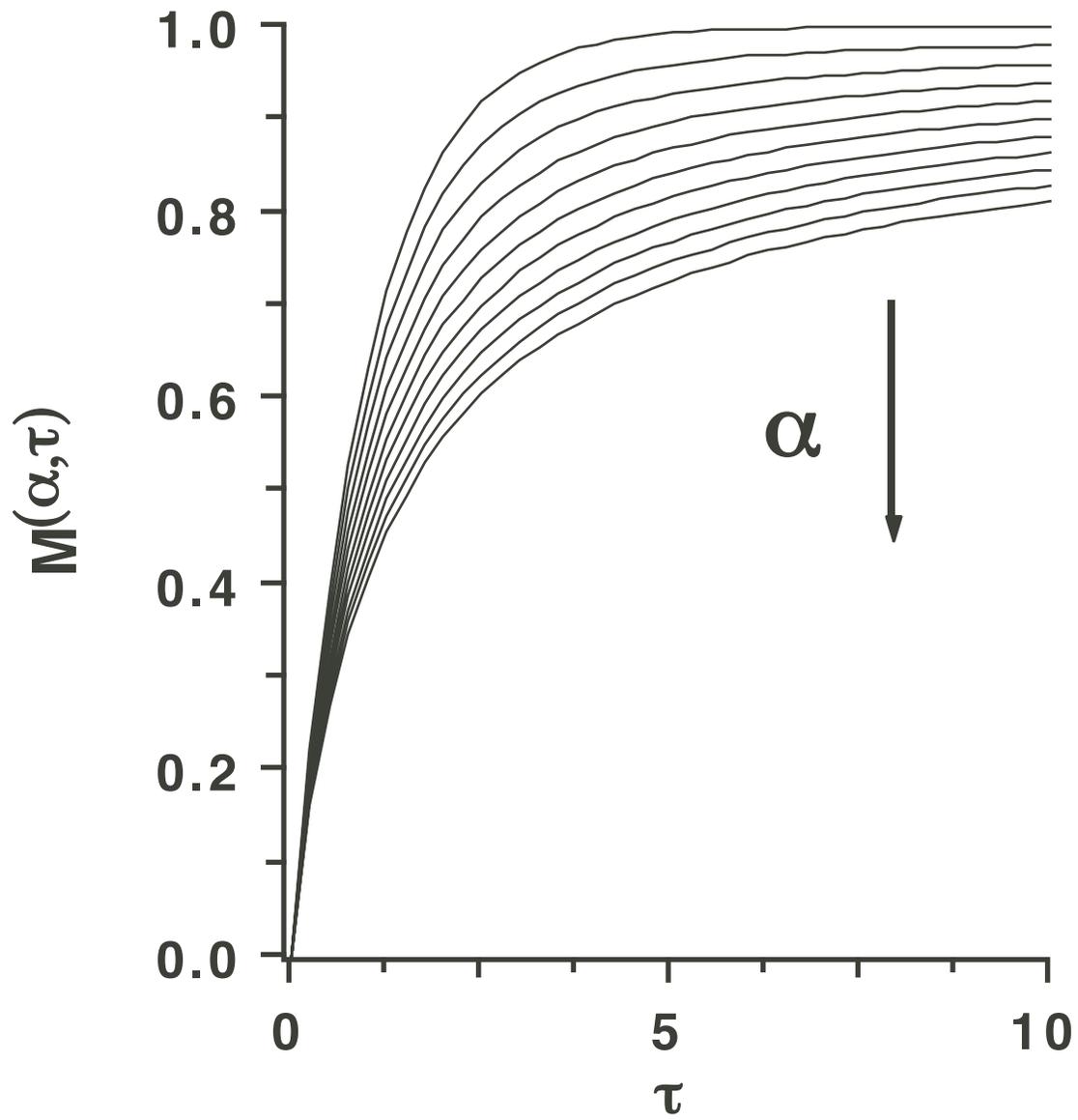

**Crowell et al; Fig. 5S**

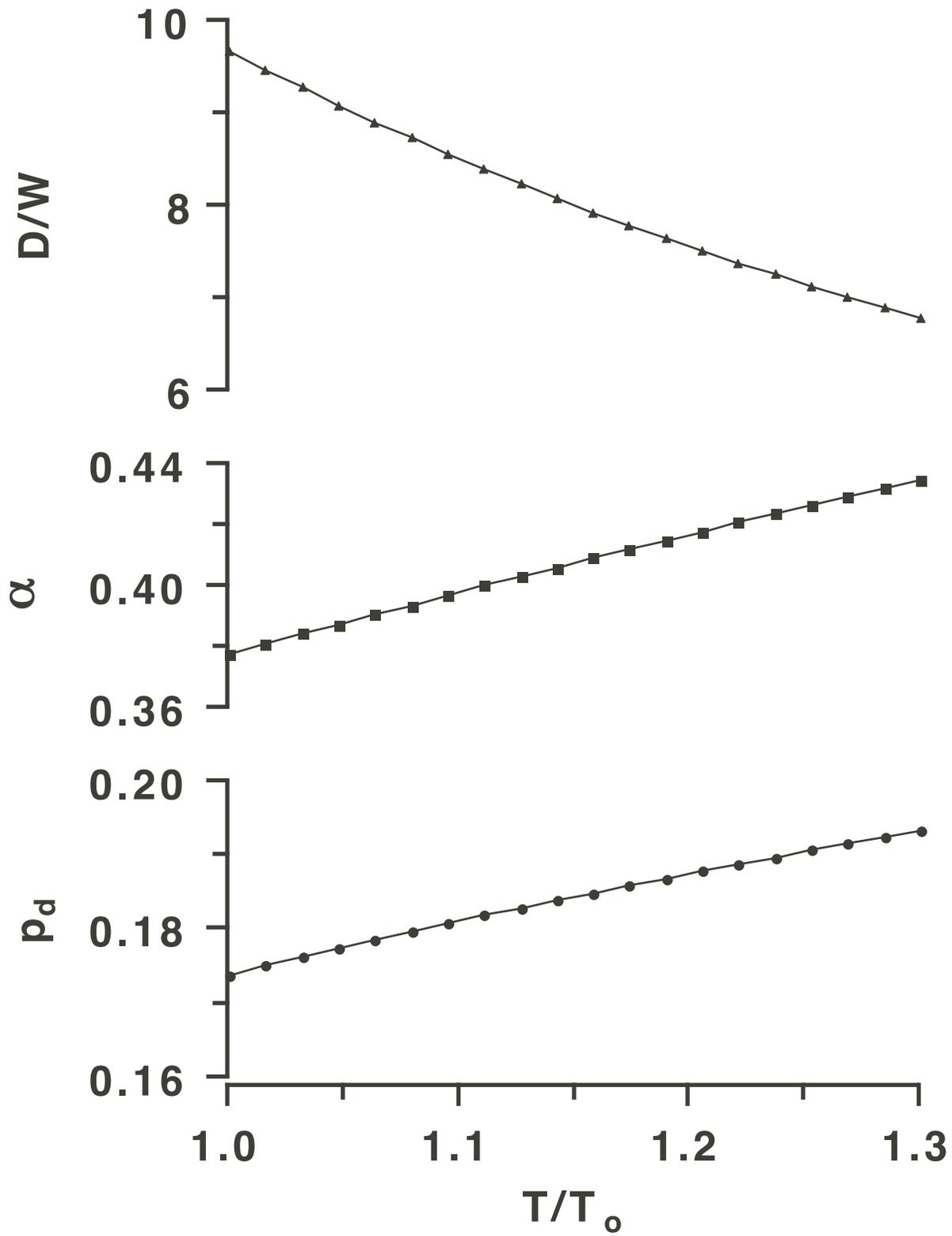

**Crowell et al; Fig. 6S**

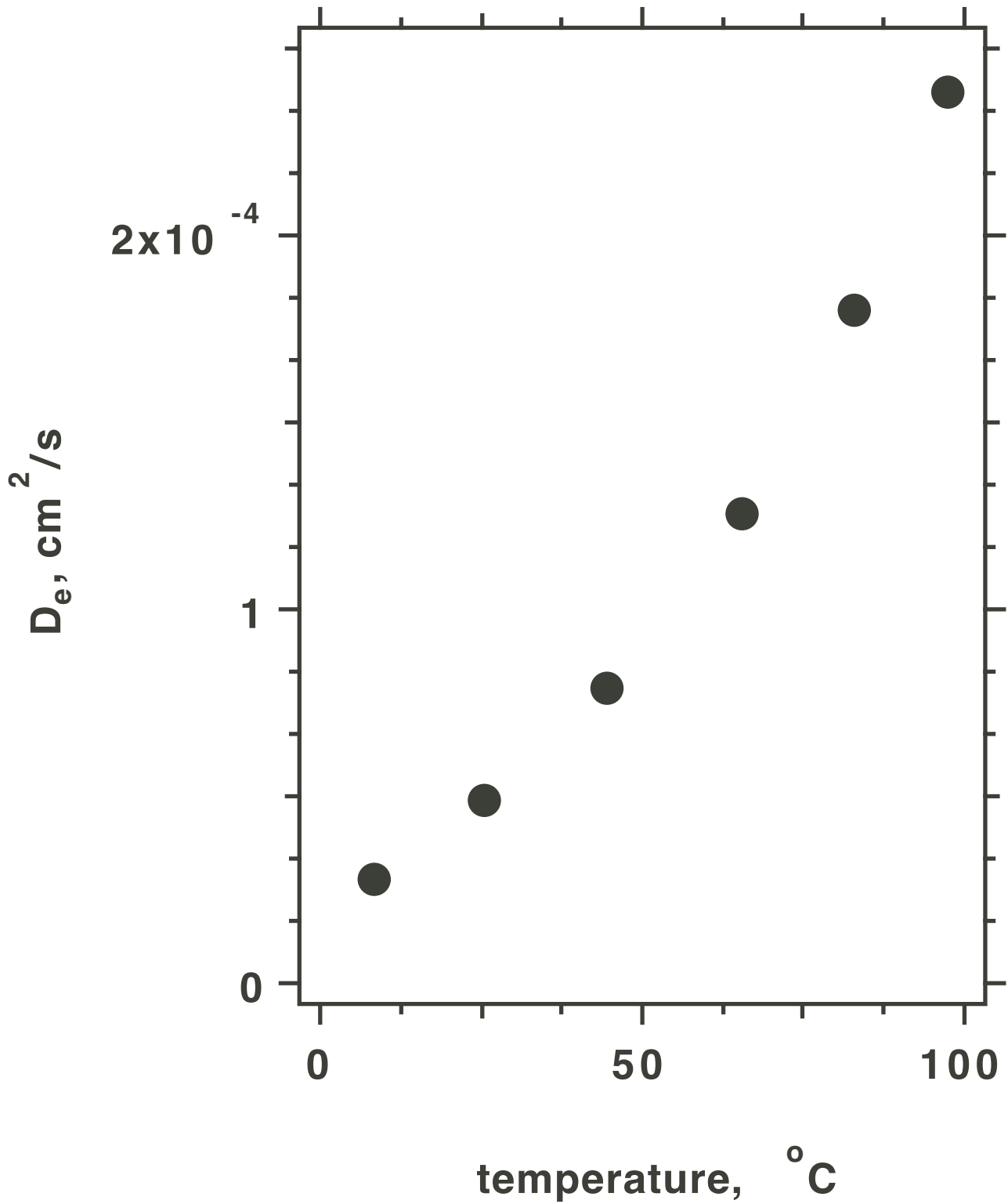

D=1.97e-5 +T*(1.05e-6 +T*(2.11e-9 +T*1.07e-10))

Crowell et al; Fig. 7S